\documentclass[onecolumn,10pt,draftcls]{IEEEtran}

\usepackage{cite}
\usepackage{tikz}
\usepackage{blindtext}
\usepackage{graphicx}
\usepackage{amsmath}
\usepackage{relsize}
\usepackage{mathrsfs}
\usepackage[capitalise]{cleveref}
\usepackage[sans]{dsfont}
\usepackage{amsbsy}
\usepackage{amssymb}
\usepackage{lipsum}
\usepackage{enumerate}
\usepackage{algorithm}
\usepackage[noend]{algpseudocode}
\usepackage{algcompatible}

\usepackage{calc}
\usepackage[framemethod=tikz,xcolor=true]{mdframed}
\newmdenv[%
    leftmargin=0.5cm,%
    roundcorner=0pt,%
    tikzsetting={draw=black, line width=2.0pt}%
    ]{SpecialText}%

\usetikzlibrary{patterns}

\DeclareMathOperator{\Equaldef}{\overset{def}{=}}
\renewcommand{\Pr}{\mathbb{P}}

\newtheorem{assumption}{Assumption}
\newtheorem{problem}{Problem}

\newtheorem{remark}{Remark}
\newtheorem{theorem}{Theorem}
\newtheorem{definition}{Definition}

\newtheorem{lemma}{Lemma}
\newtheorem{example}{Example}

\title{Optimal scheduling strategy for networked estimation with energy harvesting}
\author{Marcos M. Vasconcelos, Mukul Gagrani, Ashutosh Nayyar and Urbashi Mitra\thanks{M. M. Vasconcelos, M. Gagrani, A. Nayyar and U. Mitra are with the Ming Hsieh Department of Electrical Engineering, University of Southern California, Los Angeles, CA 90089 USA. E-mails: \texttt{\{mvasconc,mgagrani,ashutoshn,ubli\}@usc.edu}. The work of M. M. Vasconcelos and U. Mitra was supported in part by the following grants: ONR N00014-15-1-2550, NSF CCF-1718560, NSF CCF-1410009, NSF CPS-1446901, NSF CCF-1817200 and ARO W911NF1910269.}}

\begin{document}

\maketitle







\begin{abstract}
Joint optimization of scheduling and estimation policies is considered for a system with two sensors and two non-collocated estimators. Each sensor produces an independent and identically distributed sequence of random variables, and each estimator forms estimates of the corresponding sequence with respect to the mean-squared error sense. The data generated by the sensors is transmitted to the corresponding estimators, over a bandwidth constrained wireless network that can support a single packet per time slot. The access to the limited communication resources is determined by a scheduler who decides which sensor measurement to transmit based on both observations. The scheduler has an energy-harvesting battery of limited capacity, which couples the decision-making problem in time. Despite the overall lack of convexity of the team decision problem, it is shown that this system admits a globally optimal scheduling and estimation strategies under the assumption that the distributions of the random variables at the sensors are symmetric and unimodal. Additionally, the optimal scheduling policy has a structure characterized by a threshold function that depends on the time index and energy level. A recursive algorithm for threshold computation is provided.
\end{abstract}


\section{Introduction}

Reliable real-time wireless networking is an important requirement of modern cyber-physical and networked control systems \cite{Kim:2012,Bemporad:2010}. Due to their large scale, these systems are typically formed by multiple physically distributed subsystems, that communicate over a wireless network of limited capacity. One way to model this communication constraint is to assume that, at any time instant, only one packet can be reliably transmitted over the network to its destination. This constraint forces the system designer to use strategies that allocate the shared communication resources among multiple transmitting nodes. In addition to degrading the performance of the overall system, the fact that the communication among the different agents in cyber-physical systems is imperfect often leads to team-decision problems with nonclassical information structures. Such problems are usually non-convex, and are, in general, difficult to solve.

We consider a sequential remote estimation problem over a finite time horizon with non-collocated sensors and estimators. The system is comprised of multiple sensors, each of which has a stochastic process associated with it. Each sensor is paired with an estimator, which is interested in forming real-time estimates of its corresponding source process. The sensors communicate with their estimators via a shared communication network. However, at most one of the sensor's observations can be transmitted at each time due to the limited capacity of the network. In order to avoid collisions \cite{Vasconcelos:2017a,Vasconcelos:2019}, the communication is mediated by a scheduler, acting as a network manager, who observes the realization of each source and decides at each time which one, if any, gets transmitted over the communication network. In addition to the communication constraint, the framework also assumes that the scheduler operates under an energy constraint by means of a finite battery, which is capable of harvesting additional energy from the environment.

The designer's goal is to find scheduling and estimation strategies that jointly minimize an objective functional consisting of a mean-squared estimation error criterion and a communication cost. This is a team-decision problem with a non-classical information structure for which obtaining globally optimal solutions is a challenging task, in general \cite{Yuksel:2013}. However, under certain assumptions on the underlying probabilistic model, in spite of the difficulties imposed by lack of convexity this problem admits an explicit globally optimal solution, whose derivation is the  the centerpiece of this article.

\begin{figure}[t!]
\begin{center}
\includegraphics[width=0.475\textwidth]{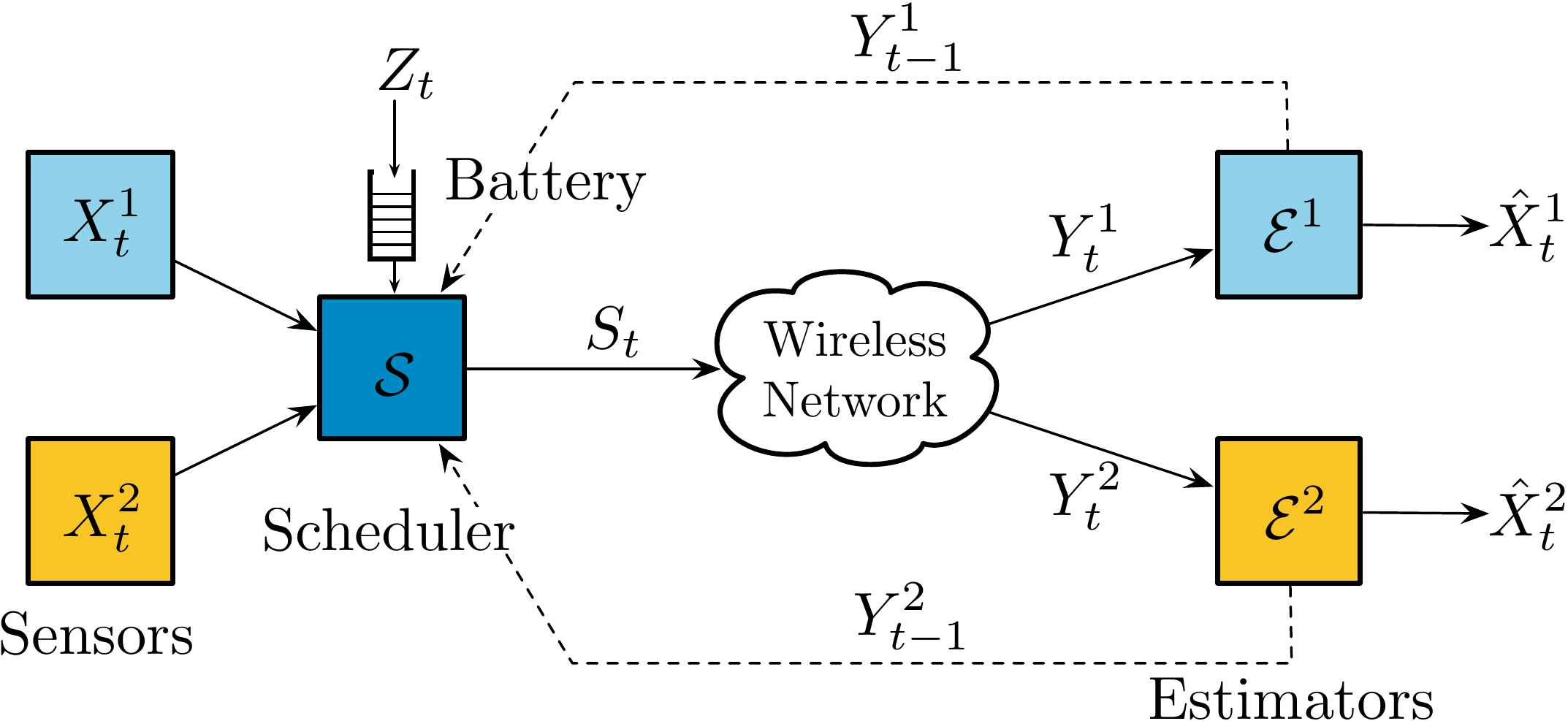}  
\caption{Schematic diagram for the remote sensing system two sensor-estimator pairs with an energy harvesting scheduler.} 
\label{fig:diagram}
\end{center}                                
\end{figure}

This problem is also motivated by applications such as the Internet of Things (IoT), where there exists a necessity to coordinate access to limited communication resources by multiple heterogeneous devices in real-time. In addition to that, in IoT applications, the network is expected to be able to support a massive number of devices for which the traditional scheduling techniques based on random access, collision resolution and retransmission are not feasibly implementable. Therefore, new scheduling schemes where decisions are driven by data such as the one proposed herein are becoming increasingly more relevant. This framework is also applicable to Wireless Body Area Networks, which are systems where multiple biometric sensors mounted on humans communicate with remote sensing stations over a wireless network \cite{Mitra:2012,Zois:2013,Zois:2016}. In order to coordinate the access of the network among multiple sensors, a mobile phone is used as a hub, collecting data and choosing in real-time which one of the measurements is transmismitted over the network, thereby acting as a scheduler. 

\subsection{Related literature}

Sequential remote estimation with a single sensor and estimator has been a well-studied problem \cite{Lipsa:2011,Nayyar:2013,Leong:2018}, for which jointly optimal sensor scheduling and estimation strategies have been derived under different structural assumptions. We are interested in problems where multiple sensors and estimators share the communication network. Systems with multiple control loops sharing a common communication network arise frequently in networked control and estimation. The authors of \cite{Walsh:2002} proposed a scheduling protocol for a networked control system with multiple sensors and actuators sharing a common communication network and analyzed its stability. The performance of event-triggered control loops closed over a shared communication network was studied in \cite{Cervin:2008} under different medium access control (MAC) protocols. The performance of an estimation problem using a contention-based MAC schemes, where each sensor can listen to the channel, before it communicates, was studied in \cite{Rabi:2010}. The design of a wireless control system using random access was considered in \cite{Gatsis:2018}, where communication policies at the sensors are designed to guarantee control performance by mitigating the effect of packet collisions. The concept of a \textit{scheduler} (or \textit{network manager}) that observes the state of multiple control loops and mediates access to the network was introduced in \cite{Henriksson:2015} and \cite{Molin:2014}. The problem of joint scheduling and remote estimation of two random variables with a single estimator was considered in \cite{Vasconcelos:2019b}, where person-by-person optimal solutions were obtained for the independent and symmetrically correlated Gaussian cases.

The problem of state estimation over a shared communication medium with multiple plants and estimators was considered in \cite{Xia:2017}.
Our problem setup is similar to the one considered in \cite{Xia:2017}. However, the observations of each sensor in \cite{Xia:2017} are Gauss-Markov processes and the performance metric is a long term average cost. More importantly, \cite{Xia:2017} fixed the estimation strategy \textit{a priori} and compared the performance of specific scheduling strategies (one of them consists in transmitting the state of the plant with the largest magnitude).

In a preliminary effort to obtain jointly optimal policies for the problem setup in \cite{Xia:2017}, the work in  \cite{Vasconcelos:2017c} considered a one-shot problem of networked estimation and characterized jointly optimal scheduling and estimation strategies under certain assumptions on the probabilistic model of the sources. In this paper, we extend the optimality result in \cite{Vasconcelos:2017c} to the sequential case over a finite horizon, when the state process of each sensor is independent of each other and is an independent and identically distributed (i.i.d.) process. However, there is a coupling of the decision-making problem accross the multiple stages due to the presence of an energy-harvesting battery of limited capacity. Our sequential problem is a team decision problem with non-classical information structure. Although such team problems are difficult to solve in general, under the condition that the sensors obervations are symmetric and unimodal probability density functions, we obtain a pair of jointly optimal scheduling and estimation strategies.

There exists an extensive literature on energy-harvesting transmitters in communications. This class of communication problems was introduced in \cite{Ozel:2010} and captures a communication feature present in several mobile systems. However, the emphasis in that line of work is in maximizing information rates. The goal of this and related papers is on minimizing a combination of estimation error and communication cost, which indirectly affects the communication rate. In remote estimation, the energy-harvesting sensors have been considered in \cite{Nayyar:2013,Ozel:2016,Leong:2018}. Related work in the field of energy-harvesting communications include \cite{Yang:2012,Michelusi:2013a,Michelusi:2013b,Michelusi:2015}. A recent survey of energy harvesting in communication and remote estimation can be found in \cite{Jog:2019}.

\subsection{Contributions}

The main contributions of this work are:

\begin{itemize}

  \item We establish the joint optimality of a pair of scheduling estimation strategies for a sequential problem formulation with i.i.d. sources and an energy-harvesting scheduler under symmetry and unimodality assumptions of the observations' pdfs.

  \item We provide a proof strategy that uses a combination of expansion of information structures and the common information approach. We show that the optimal solution of the relaxed problem also solves the original problem, and therefore, it is optimal.

  \item We illustrate the our theoretical results with numerical examples. 
\end{itemize}

\subsection{Organization}

This paper is organized into 10 sections, including the introduction. We provide the preliminary definitions for the problem formulation in Section II, and the main results in Section III. We define a relaxation of the original problem and use the common information approach to write an equivalent POMDP in Sections IV and V, respectively. The solution to the coordinator's POMDP is derived in Section VI. A numerical procedure for computing the optimal scheduling policies is provided in Section VII, and examples are provided in Section VIII. Our results are extended to an arbitrary number of sensors and to the case of unequal weights and communication costs in Section IX. We conclude in Section X, where we also point out open research directions.

\subsection{Notation}
We adopt the following notation: random variables and random vectors are represented using upper case letters, such as $X$. Realizations of random variables and random vectors are represented by the corresponding lower case letter, such as $x$. We use $X_{a:b}$ to denote the collection of random variables $(X_a,X_{a+1},\cdots,X_b)$.  The probability density function (pdf) of a continuous random variable $X$, provided that it is well defined, is denoted by $\pi$. When a random variable is distributed according to a pdf $\pi$ Functions and functionals are denoted using calligraphic letters such as $\mathcal{F}$. We use $\mathcal{N}(m,\sigma^2)$ to represent the Gaussian probability distribution of mean $m$ and variance $\sigma^2$, respectively. The real line is denoted by $\mathbb{R}$. The set of natural numbers is denoted by $\mathbb{N}$. The set of nonnegative integers is denoted by $\mathbb{Z}_{\geq 0}$. Sets are represented in blackboard bold font, such as $\mathbb{A}$. The probability of an event $\mathfrak{E}$ is denoted by $\Pr(\mathfrak{E})$; the expectation of a random variable $Z$ is denoted by $\mathbb{E}[Z]$. The indicator function of a statement $\mathfrak{S}$ is defined as follows: 
\begin{equation}
\mathbb{I}\big(\mathfrak{S}\big) \Equaldef \begin{cases}
1 & \text{if} \ \ \mathfrak{S}\ \  \text{is true}\\
0 & \text{otherwise}.
\end{cases}
\end{equation}
We also adopt the following convention:
\begin{itemize}
\item Consider the set $\mathbb{W}\Equaldef \{1, 2, \cdots, N\}$ and a function ${\mathcal{F}: \mathbb{W} \rightarrow \mathbb{R}}$ are given. If $\overline{\mathbb{W}}$ is the subset of elements that maximize $\mathcal{F}$ then $\arg \max_{\alpha \in \mathbb{W}} \mathcal{F}(\alpha)$ is defined as the smallest number in $\overline{\mathbb{W}}$.
\end{itemize}

\section{Problem statement}\label{sec:prob_formulation}

\subsection{Basic definitions}\label{sec:basic}

Consider a system with two sensor-estimator pairs and one energy harvesting scheduler. All the subsequent results hold for an arbitrary number of sensor-estimator pairs, a fact that will be formally stated in \cref{sec:N_sensors}. Therefore, the focus on two sensor-estimator pairs is without loss of generality. 

The system operates sequentially over a finite time horizon $T \in \mathbb{N}$. The role of the scheduler is to mediate the communication between the sensors and estimators such that, at any given time step, at most one sensor-estimator pair is allowed to communicate. 
We proceed to define the stochastic processes observed at the sensors. Let $X^i_t \in \mathbb{R}^{n_i}$ denote the random vector observed at the $i$-th sensor, $t\in\{1,\cdots,T\}$, $i\in\{1,2\}$. Let $n_1+n_2 = n$. We shall refer to $X^i_t$, $i\in\{1,2\}$, as outputs of information sources at time $t$.
Throughout the paper we assume that the sources are independent and identically distributed in time. Moreover, the random variables $X^i_t$ admit a pdf $\pi_i$ for all $i\in\{1,2\}$ and $t\in\{1,\cdots,T\}$.
We assume that the stochastic processes $\{ X^1_t, t\geq 1\}$ and $\{ X^2_t, t\geq 1\}$ are independent.

The scheduler operates with a battery of finite capacity denoted by $B\in \mathbb{N}$ such that $B<T$. Let the state of the battery, $E_t$, be defined as the number of energy units available at time step $t$. At each time $t$, the scheduler makes a decision $U_t \in \{0,1,2\}$, where $U_t = 0$ denotes that no transmissions are scheduled; $U_t=1$ denotes that the scheduler transmits $X^1_t$; and $U_t=2$ denotes that the scheduler transmits $X^2_t$. Each transmission depletes the battery by one energy unit and only no transmissions can be scheduled if the battery is empty, i.e., if $E_t=0$. Thus, the scheduling decision $U_t\in \mathbb{U}(E_t)$, where:
\begin{equation}\label{eq:action_set}
\mathbb{U}(E_t) \Equaldef \begin{cases}
\{0,1,2\} & \text{if} \ \ E_t >0\\
\{0\} & \text{if} \ \ E_t =0.
\end{cases}
\end{equation}

At time $t$, the scheduler harvests $Z_t$ units of energy from the environment. The random variable $Z_t$ is i.i.d. in time according to a probability mass function $p_Z(z), z\in \mathbb{Z}_{\geq 0}$,
 and is independent of the information source processes. The state of the battery evolves according to the following equation:
\begin{equation}\label{eq:c2}
E_{t+1}=\mathcal{F}(E_t,U_t,Z_t), \ \  t\in \{1,\cdots,T-1\},
\end{equation}
where
\begin{equation}\label{eq:battery}
\mathcal{F}(E_t,U_t,Z_t) \Equaldef \min\big\{ E_{t} - \mathbb{I}\big(U_t \neq 0 \big) + Z_t,B\big\},
\end{equation}
and initial energy $E_1=B$.

We will assume that the communication between the scheduler and the estimators occur over a so-called \textit{unicast network}, where only the intended estimator receives the transmitted packet. For $i\in \{1,2\}$, the observation of the estimator $\mathcal{E}^i$ at time $t$ is denoted by $Y^i_t$, which is determined according to $Y^i_t = h^i(X^i_t, U_t)$, where:
\begin{equation}\label{eq:c3}
h^i(X^i_t, U_t) \Equaldef \begin{cases}
X^i_t & \text{if} \ \ U_t = i \\
\varnothing & \text{if} \ \ U_t \neq i. 
\end{cases}
\end{equation}

\begin{remark}
One way to think about the unicast network model is that there are independent point-to-point links between different sensor and estimator pairs. At each time instant the scheduler chooses at most one of these links to be active and the others remain idle. Unicast is one of the modes of operation in the current version of the internet protocol, IPv6.
\end{remark}

\subsection{Information and strategies}

Let $\mathbf{X}_t\Equaldef (X^1_t,X^2_t)$ and $\mathbf{Y}_t\Equaldef (Y^1_t,Y^2_t)$. The scheduler decides what to transmit based on its available information at time $t$, which is $\mathcal{I}^{\mathcal{S}}_t\Equaldef\{\mathbf{X}_{1:t}, E_{1:t},\mathbf{Y}_{1:t-1}\}$. The decision variable $U_t$ is computed according to a function $f_t$ as follows:
\begin{equation}\label{eq:c4}
U_t = f_t(\mathbf{X}_{1:t}, E_{1:t},\mathbf{Y}_{1:t-1}).
\end{equation}
We refer to the collection $\mathbf{f}\Equaldef \{f_1,\cdots,f_T\}$ as the \textit{scheduling strategy} of the network manager.

Let $i\in \{1,2\}$. The estimator $\mathcal{E}^i$ computes the state estimate based on the entire history of its observations, $\mathcal{I}_t^{\mathcal{E}^i}\Equaldef\{Y_{1:t}^i\}$, according to a function $g^i_t$ as follows:
\begin{equation}\label{eq:c5}
\hat{X}^i_t = g^i_t(Y^i_{1:t}).
\end{equation}
We refer to the collection $\mathbf{g}^i
\Equaldef \{g^i_1, \cdots, g^i_T\}$ as the \textit{estimation strategy} of estimator $\mathcal{E}^i$. 

\begin{remark}
From now on, we assume that $f_t$, $g^1_t$ and $g^2_t$, $t\in\{1,\cdots,T\}$, are measurable functions with respect to the appropriate  sigma-algebras.
\end{remark}

\subsection{Cost}

We consider a performance index which penalizes the mean squared estimation error and a communication cost for every transmission made by the scheduler. 
The cost functional and optimization problem are defined as follows:
\begin{equation}\label{eq:cost}
\mathcal{J}\big(\mathbf{f},\mathbf{g}^1,\mathbf{g}^2\big) \Equaldef \sum_{t=1}^T \mathbb{E}\Bigg[  \sum_{i\in\{1,2\}}\|X^i_t-\hat{X}^i_t \|^2  +c\mathbb{I}(U_t\neq 0) \Bigg].
\end{equation}

\begin{problem}\label{prob:main}
For the model described in this section, given the statistics of the sensor's observations, the statistics of the energy-harvesting process, the battery storage limit $B$, communication cost $c$, and the horizon $T$, find scheduling and estimation strategies $\mathbf{f},\mathbf{g}^1$ and $\mathbf{g}^2$ that jointly minimize the cost $\mathcal{J}(\mathbf{f},\mathbf{g}^1,\mathbf{g}^2)$ in \cref{eq:cost}.
\end{problem}

\subsection{Signaling}

In problems of decentralized control and estimation with non-classical information structures, the optimal solutions typically involve a form of implicit communication known as \emph{signaling}. Signaling is the effect of conveying information through actions \cite{Ho:1978}, and it is the reason why problems within this class are difficult to solve, e.g. \cite{Witsenhausen:1968}.

\begin{figure}[!t]
    \begin{center}
    \includegraphics[scale=0.4]{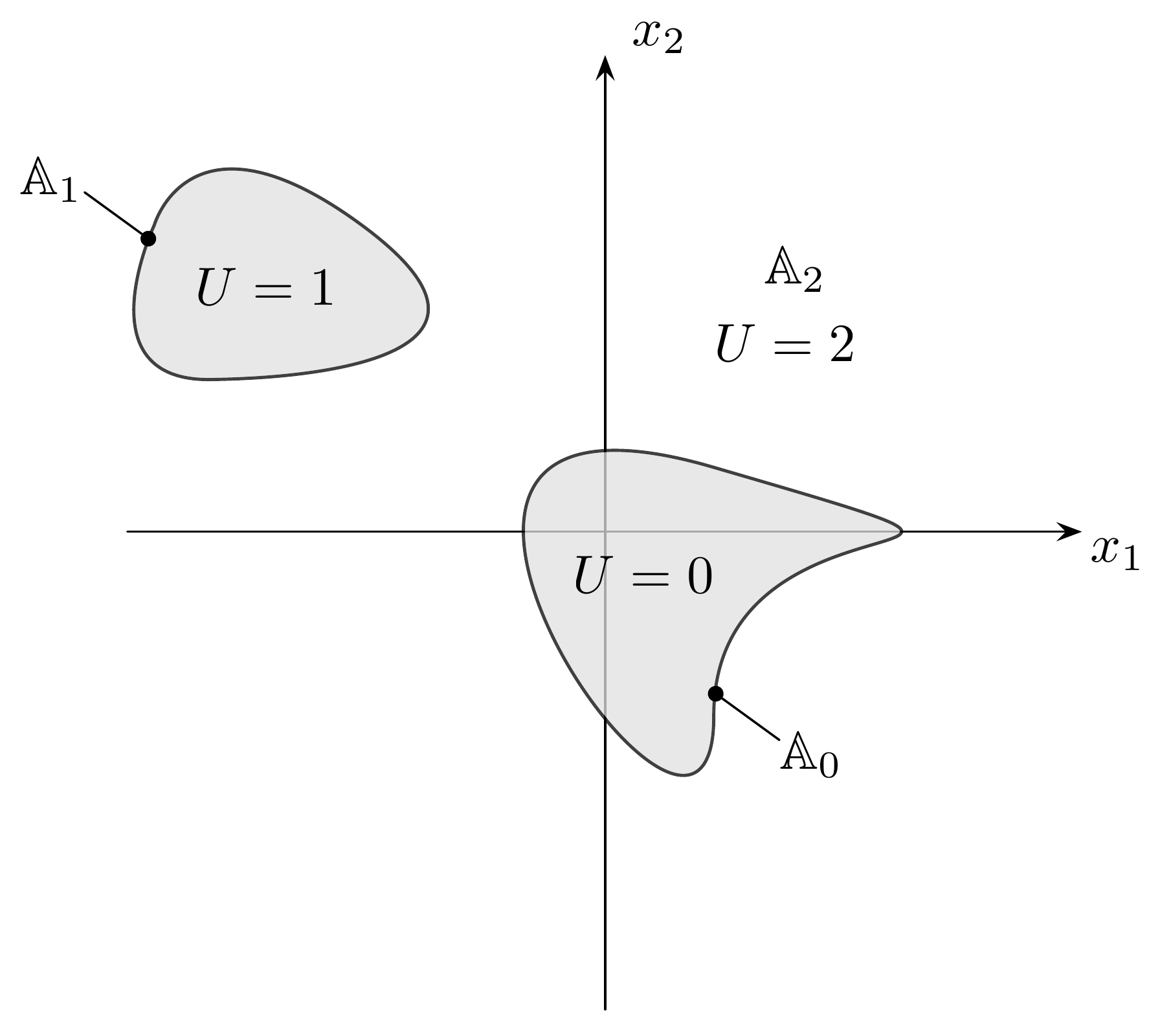}
\end{center}
\caption{Partition of the observation space used to illustrate the issue of signaling in problems of networked estimation.}
\label{fig:signaling}
\end{figure}

In order to illustrate the fundamental difficulty imposed by signaling, consider an instance of \cref{prob:main} with two zero-mean independent scalar sources and $T=1$. Assume that the scheduler makes its decision according to the partition of the observation space shown in \cref{fig:signaling}, where $\mathbb{A}_0\cup\mathbb{A}_1\cup\mathbb{A}_2=\mathbb{R}^2$ and $(x_1,x_2)\in\mathbb{A}_i$ implies that $U=i$, $i\in\{0,1,2\}$. Suppose that the scheduler observes $(x_1,x_2) \in \mathbb{A}_1$, which implies that $U=1$ and consequently, $Y_1 = x_1$ and $Y_2 = \varnothing$. The optimal estimate used by $\mathcal{E}_2$ in this case is  
\begin{equation}
\begin{aligned}
 \hat{X}_2  = & \ \mathbb{E}[X_2 \mid Y_2 = \varnothing]\\
  =  & \ \mathbb{E}[X_2 \mid (X_1,X_2) \in \mathbb{A}_2^{c}],
\end{aligned}
\end{equation}
which may correspond to a different numerical value than if we simply used a naive estimate $\mathbb{E}[X_2]=0$.

Therefore, the optimal estimation strategy depends on the scheduling strategy being used. This coupling between scheduling and estimation strategies is what makes \cref{prob:main} nontrivial even when $T=1$.

\section{Main result}

The following definition will be used to state our main result.

\begin{definition}[Symmetric and unimodal probability density functions]
Let $\pi: \mathbb{R}^{n} \rightarrow \mathbb{R}$ be a probability density function (pdf). The pdf $\pi$ is symmetric and unimodal around $a \in \mathbb{R}^n$ if it satisfies the following property:
\begin{equation}
\|x - a\| \leq \|y - a\| \Rightarrow \pi(x) \geq \pi(y), \ \  x,y \in \mathbb{R}^n. 
\end{equation}
\end{definition}



 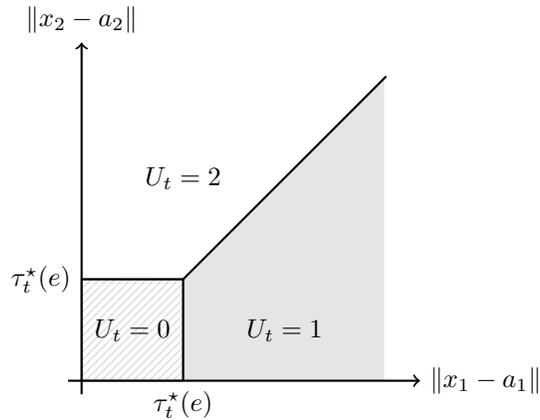
\begin{figure}[!t]
\centering
\begin{tikzpicture}[scale=0.9]

\draw[ thick,pattern=north east lines,opacity=0.4](0,0)--(0,1.5)--(1.5,1.5)--(1.5,0)--(0,0);

\fill[lightgray,opacity=0.4] (1.5,1.5) -- (4.47,4.5) -- (4.47,0) --(1.5,0);


\draw[ thick] (0,1.5)--(1.5,1.5)--(1.5,0);

\draw[ thick] (1.5,1.5)--(4.5,4.5);


\draw[->, thick] (-0.2,0)--(5,0) node[right]{$\|x_1-a_1\|$};
\draw[->, thick] (0,-0.2)--(0,5) node[above]{$\|x_2-a_2\|$};

\node at (-0.6,1.5) { $\tau^{\star}_t(e)$};
\node at (1.5,-0.35) { $\tau^{\star}_t(e)$};

\node at (0.75,0.75) {$U_t=0$};

\node at (3,0.75) { $U_t=1$};

\node at (1.5,3) { $U_t=2$};


\end{tikzpicture}
\label{fig:partition}
\caption{Pictorial representation of the optimal scheduling $f^{\star}_t(\mathbf{x},e)$ indexed by the function $\tau^{\star}_t$.}
\end{figure}



\begin{theorem}\label{thm:main}
Provided that $\pi_1$ and $\pi_2$ are symmetric and unimodal around $a^1 \in \mathbb{R}^{n_1}$ and $a^2 \in \mathbb{R}^{n_2}$, respectively, the strategy profile $\big(\mathbf{f}^{\star}, \mathbf{g}^{1\star}, \mathbf{g}^{2\star}\big)$ is globally optimal for \cref{prob:main}, where $\mathbf{f}^{\star}$ is a vector whose $t$-th component is defined as:
\begin{equation}
f^\star_t(\mathbf{x},e) \Equaldef \begin{cases} 
0,  \ \  \text{if} \ \ \underset{i\in\{1,2\}}{\max} \big\{\|x^i-a^i\|\big\} \leq \tau^\star_t(e) \\
\arg \underset{i\in\{1,2\}}{\max}  \|x^i - a^i\|, \ \ \ \ \   \text{otherwise,}
\end{cases}
\end{equation}
where $\tau^\star_t: \mathbb{Z} \rightarrow  \mathbb{R}$; and $\mathbf{g}^{i\star}$ is a vector whose $t$-th component is defined as:
\begin{equation}
g^{i\star}_t\big(y^i) \Equaldef \begin{cases} 
x^i & \textup{if} \ \ y^i = x^i \\
a^i & \textup{if} \ \ y^i = \varnothing.
\end{cases}
\end{equation}
\end{theorem}

\section{Information Structures}

Problem \ref{prob:main} can be understood as a sequential stochastic team with three decision makers: the scheduler and the two estimators. One key aspect to note is that Problem \ref{prob:main} has a non-classical information structure. Such team problems are usually non-convex and their solutions are found on a case-by-case basis. Our analysis relies on the \textit{common information approach} \cite{Nayyar:2013b}, where the idea is to transform the decentralized problem into an equivalent centralized one where the information for decision-making is the common information among all the decision makers in the decentralized system. 

We begin by establishing a structural result for the optimal scheduling strategy. The following lemma states that the scheduler may ignore the past state observations at each sensor and the past states of the battery without any loss of optimality.

\begin{lemma}\label{lem:scheduler_struct}
Without loss of optimality, the scheduler can be  restricted to strategies of the form:
\begin{equation}
U_t =f_t \big(\mathbf{X}_{t},E_{1:t},\mathbf{Y}_{1:t-1} \big).
\end{equation}
\end{lemma}

\begin{IEEEproof}
Let the strategy profile of the estimators $\mathbf{g}^1$ and $\mathbf{g}^2$ be arbitrarily fixed. The problem of selecting the best scheduling policy (for the fixed estimation strategy profiles $\mathbf{g}^1$ and $\mathbf{g}^2$) simplifies to a Markov Decision Process (MDP), whose state is defined as $S_t \Equaldef (\mathbf{X}_t,E_{1:t},\mathbf{Y}_{1:t-1})$. Using simple arguments involving conditional probabilities and the basic definitions of \cref{sec:basic}, we can show that the state process $\{S_t,t\geq 1\}$ is a controlled Markov chain, i.e., 
\begin{equation}\label{eq:Markov}
\mathbb{P}(S_{t+1} \mid S_{1:t}, U_{1:t}) = \mathbb{P}(S_{t+1} \mid S_{t}, U_{t}).
\end{equation}
%

The cost incurred at time $t$ of the equivalent MDP is: 
\begin{align*}
 \rho(S_t,U_t)  &\Equaldef \sum_{i\in\{1,2\}} \big\|X_t^i - \hat{X}_t^i\big\|^2 + c\mathbb{I}(U_t\neq 0)  \\
&\stackrel{(a)}{=} \sum_{i\in\{1,2\}}  \big\|X_t^i - g_t^i (Y_{1:t}^i) \big\|^2 + c\mathbb{I}(U_t\neq 0) \\
&\stackrel{(b)}{=}\sum_{i\in\{1,2\}}  \big\|X_t^i - g_t^i \big(Y_{1:t-1}^i,h^i(X_t^i,U_t)\big) \big\|^2  + c\mathbb{I}(U_t\neq 0), 
\end{align*}
where $(a)$ follows from \cref{eq:c5} and $(b)$ follows from \cref{eq:c3}.

Thus, the problem of finding the optimal scheduling strategy to minimize the cost $\mathcal{J}\big(\mathbf{f}, \mathbf{g}^1, \mathbf{g}^2 \big)$ becomes equivalent to finding the optimal decision strategy for an MDP with state process $S_t$ and instantaneous cost $\rho(S_t,U_t)$. Standard results for MDPs \cite{Kumar:1986} imply that there exists an optimal scheduling strategy of the form in lemma. Since this is true for any arbitrary $\mathbf{g}^1$ and $\mathbf{g}^2$, it is also true for the globally optimal  $\mathbf{g}^{1\star}$ and $\mathbf{g}^{2\star}$.
\end{IEEEproof}

Under the structural result in \cref{lem:scheduler_struct}, the information sets available at the network manager and estimators can be reduced to:
\begin{align}
\mathcal{I}_t^{\mathcal{S}} &\Equaldef \big\{\mathbf{X}_{t},E_{1:t},\mathbf{Y}_{1:t-1} \big\} \label{eq:Information_U}\\
\mathcal{I}_t^{\mathcal{E}^i} &\Equaldef \big\{Y^{i}_{1:t} \big\}, \ \ i \in\{1,2\} \label{eq:Information_E},
\end{align}
without any loss of optimality. However, the information structure described by \cref{eq:Information_U,eq:Information_E} do not share any common information. In other words, the information sets $\mathcal{I}_t^{\mathcal{S}}$, $\mathcal{I}_t^{\mathcal{E}^1}$ and $\mathcal{I}_t^{\mathcal{E}^2}$have no common random variables, a fact that limits the utility of the common information approach. We resort to a technique which consists of judiciously expanding the information available at the decision makers such that the common information approach can be more profitably employed.  

\subsection{Information structure expansion}

We expand the estimators' information sets to the following:
\begin{equation}
\bar{\mathcal{I}}_t^{\mathcal{E}^1} \Equaldef \big\{E_{1:t},\mathbf{Y}_{1:t-1},Y_t^1 \big\}\ \label{eq:estimator_expanded_1}
\end{equation}
\begin{equation}
\bar{\mathcal{I}}_t^{\mathcal{E}^2} \Equaldef \big\{E_{1:t},\mathbf{Y}_{1:t-1},Y_t^2 \big\}. \label{eq:estimator_expanded_2}
\end{equation}

The optimal cost for Problem \ref{prob:main} under an expanded information structure is at least as good as the optimal cost under the original information structure (having more information at each estimator cannot worsen its performance). Moreover, if the optimal solution under the expanded information structure is adapted to the original information structure, then this solution is also optimal under the original information structure \cite[Proposition 3.5.1]{Yuksel:2013}. 



We proceed by defining another problem identical to Problem \ref{prob:main} but with expanded information sets at the estimators. 

\begin{problem}\label{prob:expanded_info}
Consider the model of \cref{sec:prob_formulation} with the expanded information sets of \cref{eq:estimator_expanded_1,eq:estimator_expanded_2} at the estimators $\mathcal{E}^1$ and $\mathcal{E}^2$, respectively. 
Given the statistics of the sensors' observations, the statistics of the energy harvested at each time, the battery storage limit $B$, communication cost $c$, and the horizon $T$, find the scheduling and estimation strategies $ \mathbf{f}, \mathbf{g}^1$  and $\mathbf{g}^2  $ that jointly minimize the cost $\mathcal{J}\big(\mathbf{f}, \mathbf{g}^1, \mathbf{g}^2 \big)$ in \cref{eq:cost}.
\end{problem}

Under the expanded information structure, the common information among the decision makers is:
\begin{equation}
\mathcal{I}^{\mathrm{com}}_t \Equaldef \big\{E_{1:t},\mathbf{Y}_{1:t-1} \big\}.
\end{equation}
Notice that the common information contains several variables which were not originally available to the estimators. However, we will show that the optimal estimation strategy for \cref{prob:expanded_info} does not depend on this additional information.

The following lemma provides a structural result for the estimation strategies under the expanded information sets.

\begin{lemma}\label{lem:estimator_pbpo}
Without loss of optimality, the search for optimal strategies for estimator $\mathcal{E}^i$ can be restricted to functions of the form:
\begin{equation}
g_t^i(E_{1:t},\mathbf{Y}_{1:t-1}, Y^i_t) = \begin{cases} 
X_t^i & \textup{if} \ \ Y_t^i = X^i_t \\
\tilde{g}_t^i(E_{1:t},\mathbf{Y}_{1:t-1}) & \textup{otherwise}.
\end{cases}
\end{equation}
\end{lemma}

\begin{IEEEproof}
Let the strategy of the scheduler be fixed to some arbitrary $\mathbf{f}$. We can view \cref{prob:expanded_info} from the perspective of the estimator $\mathcal{E}^i$ at time $t$ as follows:
\begin{equation}
\inf_{g^i_t} \ \ \mathbb{E} \big[\|X_t^i - \hat{X}_t^i \|^2 \big] + \tilde{\mathcal{J}},
\end{equation}
where
\begin{equation}
\tilde{\mathcal{J}} \Equaldef \mathbb{E} \Bigg[ \sum_{k=1}^T c \mathbb{I}(U_k \neq 0) +\sum_{k=1}^T \sum_{j \neq i} \| X_k^j - \hat{X}_k^j \|^2 \\ + \sum_{k \neq t} \| X_k^i - \hat{X}_k^i \|^2  \Bigg].
\end{equation}

Since $g_t^i$ does not affect the term $\tilde{\mathcal{J}}$, the optimal estimate can be computed by solving:
\begin{equation}
\inf_{g^i_t} \ \ \mathbb{E} \big[\|X_t^i - \hat{X}_t^i \|^2 \big].
\end{equation}
This is the standard MMSE estimation problem whose solution is the conditional mean, i.e.,
\begin{equation}
\hat{X}_t^i = \mathbb{E}\big[X_t^i \ \big| \  \bar{\mathcal{I}}_t^{\mathcal{E}^i}\big].
\end{equation}
Therefore, the optimal estimation strategy is of the form:
\begin{align} \label{eq:estimator_pbpo}
g_t^{i\star}(\bar{\mathcal{I}}_t^{\mathcal{E}^i}) = \begin{cases} 
X_t^i & \textup{if} \ \ Y_t^i=X_t^i  \\
\mathbb{E}\big[ X_t^i \ \big| \ E_{1:t}, \mathbf{Y}_{1:t-1}, Y^i_t=\varnothing \big] & \textup{otherwise}.
\end{cases}
\end{align}
Notice that $(E_{1:t},\mathbf{Y}_{1:t-1})$ is known to $\mathcal{E}^i$ in Problem \ref{prob:expanded_info}. Thus, 
\begin{equation}
\tilde{g}_t^i(E_{1:t},\mathbf{Y}_{1:t-1}) \Equaldef  \mathbb{E}\big[ X_t^i \ \big|\ E_{1:t}, \mathbf{Y}_{1:t-1}, Y^i_t=\varnothing \big].
\end{equation}
 
 Since \cref{eq:estimator_pbpo} holds for any $\mathbf{f}$, it also holds for the globally optimal scheduling strategy $\mathbf{f^\star}$. Therefore, the optimal estimate is of the form given in the lemma. 
\end{IEEEproof}

\section{An equivalent problem with a coordinator}\label{sec:common_info}

In this section, we will formulate a problem which will be used to solve \cref{prob:expanded_info}. We consider the model of \cref{sec:prob_formulation} and introduce a fictitious decision maker referred to as the \textit{coordinator}, which has access to the common information $\mathcal{I}^{\mathrm{com}}_t$. The coordinator is the only decision maker in the new problem. The scheduler and the estimators act as ``passive decision makers'' to which strategies chosen by the coordinator are prescribed. 

The equivalent system operates as follows: At each time $t$, based on $\mathcal{I}_t^{\mathrm{com}}$, the coordinator chooses a map $\Gamma_t: \mathbb{R}^{n_1} \times \mathbb{R}^{n_2} \rightarrow \{0,1,2 \} $ for the network manager, and a vector $\tilde{X}_t^i \in \mathbb{R}^{n_i}$ for each estimator $\mathcal{E}^i$, $i \in \{1,2 \}$. The function $\Gamma_t$ and vectors $\tilde{X}_t^1$ and $\tilde{X}_t^2$ are referred to as the scheduling and estimation \textit{prescriptions}.
The scheduler uses its prescription to evaluate $U_t$ according to:
\begin{equation}
U_t = \Gamma_t(\mathbf{X}_t).
\end{equation}
The estimator $\mathcal{E}^i$ uses its prescription to compute the estimate $\hat{X}_t^i$ according to:
\begin{equation}\label{estimate_prescription}
\hat{X}_t^i = \begin{cases} 
X_t^i & \textup{if} \ \ Y_t^i = X_t^i \\
\tilde{X}_t^i & \textup{otherwise}.
\end{cases}
\end{equation}

The coordinator selects its prescriptions for the scheduler and the estimators using strategies $d_t, \ell_t^1$ and $\ell_t^2$ as follows:
\begin{equation}
\Gamma_t = d_t(E_{1:t}, \mathbf{Y}_{1:t-1}) 
\end{equation}
and
\begin{equation}
\tilde{X}_t^i = \ell_t^i (E_{1:t}, \mathbf{Y}_{1:t-1}),\ \ i \in \{1,2 \}.
\end{equation}
We refer to the collections $\mathbf{d} \Equaldef \{d_1,\cdots,d_T \}$ and $\boldsymbol{\ell}^i \Equaldef \{\ell_t^i,\cdots,\ell_T^i \}$ as the prescription strategies for the scheduler and the estimator $\mathcal{E}^i$, respectively. The strategies $\boldsymbol{\ell}^1$ and $\boldsymbol{\ell}^2$ must be a valid estimation strategies in Problem 2. The strategy $\boldsymbol{d}$ must be such that 
\begin{equation}
f_t(X_t E_{1:t}, Y_{1:t-1}) \Equaldef \big[d_t(E_{1:t},Y_{1:t-1})\big](X_t) 
\end{equation}
is a valid scheduling strategy in Problem 2. The cost incurred by the prescription strategies $\mathbf{d},\boldsymbol{\ell}^1$ and $\boldsymbol{\ell}^2$ is identical as in \cref{eq:cost}, that is,
\begin{equation}\label{eq:cost_coord}
\hat{\mathcal{J}}(\mathbf{d},\boldsymbol{\ell}^1,\boldsymbol{\ell}^2) = \sum_{t=1}^T \mathbb{E}\Bigg[ c\mathbb{I}(U_t\neq 0) + \sum_{i\in\{1,2\}}\|X^i_t-\hat{X}^i_t \|^2  \Bigg].
\end{equation}

\begin{problem}\label{prob:coord}
Find prescription strategies $\mathbf{d}, \boldsymbol{\ell}^1$, and $\boldsymbol{\ell}^2$ that jointly minimize $\hat{\mathcal{J}}(\mathbf{d},\boldsymbol{\ell}^1,\boldsymbol{\ell}^2)$.
\end{problem}

Problem \ref{prob:coord} is equivalent to Problem \ref{prob:expanded_info} in the sense that 
for every scheduling strategy $\mathbf{f}$ and estimation strategies $\mathbf{g}^1,\mathbf{g}^2$ in Problem \ref{prob:expanded_info} there exist prescription strategies $\mathbf{d}, \boldsymbol{\ell}^1$ and $\boldsymbol{\ell}^2$ such that $\mathcal{J}(\mathbf{f},\mathbf{g}^1,\mathbf{g}^2) = \hat{\mathcal{J}}(\mathbf{d},\boldsymbol{\ell}^1,\boldsymbol{\ell}^2)$ and vice-versa. Thus, solving Problem \ref{prob:coord} allows us to obtain optimal $\mathbf{f}^\star,\mathbf{g}^{1\star}$ and $\mathbf{g}^{2\star}$ for Problem \ref{prob:expanded_info}. The same technique is used in \cite{Nayyar:2013} to prove a similar equivalence in a problem involving a single sensor-estimator pair.  


Problem \ref{prob:coord} can be described as a centralized POMDP as follows:

\begin{enumerate}[($i$)]

\item \textbf{\textit{State process:}} \\
The state is $S_t \Equaldef (\mathbf{X}_t,E_t)$.

\vspace{5pt}

\item \textbf{\textit{Action process:}}\\
Let the set $\mathbb{A}(E_t)$ be defined as the collection of all measurable functions from $\mathbb{R}^{n_1} \times \mathbb{R}^{n_2} \rightarrow \mathbb{U}(E_t)$, where $\mathbb{U}$ is defined in \cref{eq:action_set}.
The coordinator selects the prescription for the network manager, $\Gamma_t \in \mathbb{A}(E_t)$, and the prescriptions for the estimators $ \tilde{X}_t^1 \in \mathbb{R}^{n_1}$ and $ \tilde{X}_t^2 \in \mathbb{R}^{n_2}$. 

\vspace{5pt}

\item \textbf{\textit{Observations:}}\\
After choosing its action at time $t$, the coordinator observes $Y_t$ and $E_{t+1}$.

\vspace{5pt}

\item \textbf{\textit{Instantaneous cost:}}\\
Let $\tilde{\mathbf{X}}_t \Equaldef (\tilde{X}^1_t,\tilde{X}^2_t)$. The instantaneous cost incurred is given by
\begin{equation}\label{eq:rho}
\rho(\mathbf{X}_t,\Gamma_t, \tilde{\mathbf{X}}_t) \Equaldef \begin{cases}
\underset{i\in\{1,2\}}{\sum}\| X_t^i - \tilde{X}_t^i \|^2  & \text{if} \ \Gamma_t (\mathbf{X}_t) = 0 \\
 c + \| X_t^2 - \tilde{X}_t^2 \|^2 & \text{if} \ \Gamma_t (\mathbf{X}_t) = 1 \\
 c + \| X_t^1 - \tilde{X}_t^1 \|^2 & \text{if } \Gamma_t (\mathbf{X}_t) = 2.
\end{cases} 
\end{equation}

\vspace{5pt}

\item \textbf{\textit{Markovian dynamics:}}
\\
Since $\mathbf{X}_t$ is an i.i.d process, $\mathbf{X}_{t+1}$ is independent of $S_t$. The evolution of the energy $E_{t+1}$ is given by:
\begin{equation}\label{eq:coord_energy_dynamics}
E_{t+1} = \min\big\{E_t - \mathbb{I}\big(\gamma_t(\mathbf{X}_t) \neq 0\big) + Z_t, B \big\}.
\end{equation}
Noticing that \cref{eq:coord_energy_dynamics} can be written as a function of the state $S_t$, action $\gamma_t$ and the noise $Z_t$, the state $S_t$ satisfies \cref{eq:Markov} and forms a controlled Markov chain.

\end{enumerate}

\subsection{Dynamic program}

Having established that Problem \ref{prob:coord} is a POMDP, the optimal prescriptions can be computed by solving a dynamic program whose information state is the belief of the state process given the common information. However, since $E_t$ is perfectly observed, the coordinator only needs to form a belief on $\mathbf{X}_t$. 
Let $\mathbf{x}= (x^1,x^2)$. We define the belief state at time $t$ as:
\begin{equation}
\Pi_{t}(\mathbf{x})\Equaldef \mathbb{P}\big(\mathbf{X}_t=\mathbf{x} \mid  E_{1:t},\mathbf{Y}_{1:t-1}\big).
\end{equation}

Since the sources are i.i.d. and independent of the energy process, we have:
\begin{equation}
\Pi_t(\mathbf{x}) = \pi(\mathbf{x}), \ \ t \in \{1, \cdots, T\},
\end{equation}
where, due to the independence of the sources,
\begin{equation}
\pi(\mathbf{x})=\pi_1(x^1)\pi_2(x^2).
\end{equation}


\begin{lemma}\label{lem:DP}
Define the functions $\mathcal{V}_t^{\pi} : \mathbb{Z} \rightarrow \mathbb{R}$ for $t \in \{0,1,\cdots, T+1\}$ as follows:

\begin{equation}\label{eq:v_T}
\mathcal{V}_{T+1}^{\pi}(e) \Equaldef 0, \ \ e \in \{0,1,\ldots,B \},
\end{equation}
and
\begin{equation}\label{eq:v_t}
\mathcal{V}_{t}^{\pi}(e) \Equaldef \inf_{\tilde{\mathbf{x}}_t,\gamma_t} \mathbb{E} \Big[ \rho(\mathbf{X}_t,\gamma_t,\tilde{\mathbf{x}}_t)  + \mathcal{V}_{t+1}^{\pi}\big(\mathcal{F}\big(e, \gamma_t(\mathbf{X}_t), Z_t\big)\big) \Big], 
\end{equation}
where $\tilde{\mathbf{x}}_t \in \mathbb{R}^n$, $\gamma_t \in \mathbb{A}(e)$.

If the infimum in \cref{eq:v_t} is achieved, then at each time $t \in \{1,\cdots,T\}$ and for each $e \in \{0,1,\cdots,B \}$, the minimizing $\gamma_t$  and  $\tilde{\mathbf{x}}_t$ in \cref{eq:v_t} determines the optimal prescriptions for the network manager and the estimators, respectively. Furthermore, $\mathcal{V}_1(B)$ is the optimal cost for Problem \ref{prob:coord}.
\end{lemma}

\begin{IEEEproof}
This result follows from standard dynamic programming arguments for POMDPs.
\end{IEEEproof}

\section{Solving the dynamic program}

In this section, we will find the optimal prescriptions using the dynamic program in \cref{lem:DP}. For the remainder of this section, we will assume that $\pi_1$ and $\pi_2$ are symmetric and unimodal around $0$. The same arguments apply for general $a^i \in \mathbb{R}^{n_i}$, $i \in\{1,2\}$. 

Note that each step of the dynamic program in \cref{eq:v_t} is an optimization problem with respect to $\mathbf{\tilde{x}}_t$ and $\gamma_t$. This is an infinite-dimensional optimization problem since $\gamma_t$ is a mapping which lies in $\mathbb{A}(E_t)$. The next lemma will describe the structure of the optimal prescription for the scheduler and show that the infinite dimensional optimization in \cref{eq:v_t} can be reduced to a finite dimensional problem with respect to the vector $\mathbf{\tilde{x}}_t$. For that purpose, we define the functions $\mathcal{C}_{t+1}^0,\mathcal{C}_{t+1}^1: \mathbb{Z}\rightarrow \mathbb{R}$ as follows:
\begin{align}
\mathcal{C}_{t+1}^{0}(e) & \Equaldef  \mathbb{E}\Big[\mathcal{V}^{\pi}_{t+1}\big(\min\{e+Z_t,B\}\big)\Big] \label{eq:auxiliary1}\\
\mathcal{C}_{t+1}^{1}(e) & \Equaldef  c+\mathbb{E}\Big[\mathcal{V}^{\pi}_{t+1}\big(\min\{e-1+Z_t,B\}\big)\Big]\label{eq:auxiliary2}.
\end{align}

\begin{lemma}\label{lem:finite_dimensional}
Suppose the prescription to the estimators are $\tilde{x}_t^1, \tilde{x}_t^2$ at time $t$. Then, the optimal prescription to the scheduler has the following form when $e>0$:
\begin{equation}
\gamma_t^{\star}(\mathbf{x}_t) \Equaldef \begin{cases} 
0,  \ \ \text{if} \ \ \underset{i\in\{1,2\}}{\max} \big\{\|x_t^i - \tilde{x}_t^i \| \big\} \leq \tau^\star_t(e) \\
\arg \underset{i\in\{1,2\}}{\max}\big\{\|x_t^i - \tilde{x}_t^i \|\big\}, \ \ \text{otherwise},
\end{cases}
\end{equation}
where $\tau^\star_{t} (e) \Equaldef \sqrt{\mathcal{C}_{t+1}^1(e) - \mathcal{C}_{t+1}^0(e)}$\footnote{The function $\mathcal{C}^1_{t+1}(e)$ is larger than $\mathcal{C}^0_{t+1}(e)$. Therefore, the threshold $\tau^{\star}_t(e)$ is a real number for all $e\in\{1,\cdots,B\}$ and $t\in\{1,\cdots,T\}$.}. Moreover, the value function $\mathcal{V}_t^{\pi}$ of \cref{lem:DP} can be obtained by solving the finite dimensional optimization in \cref{eq:finite_dimensional_opt}.

\begin{equation}\label{eq:finite_dimensional_opt}
\mathcal{V}_{t}^{\pi}(e) =  \begin{cases}
 \underset{\tilde{\mathbf{x}}_t}{\inf} \ \mathbb{E} \Big[ \underset{i\in\{1,2\}}{\mathlarger{\sum}}\|X_t^i -  \tilde{x}_t^i\|^2 \Big] + \mathcal{C}_{t+1}^0(e)  & \text{if} \ \ e=0 \vspace{10pt}
\\
 \underset{\tilde{\mathbf{x}}_t}{\inf} \ \mathbb{E} \Big[ \min\Big\{ \underset{i\in\{1,2\}}{\mathlarger{\sum}}\|X_t^i -  \tilde{x}_t^i\|^2 + \mathcal{C}^0_{t+1}(e), \|X^2_t - \tilde{x}^2_t \|^2 + \mathcal{C}^1_{t+1}(e),
\|X^1_t - \tilde{x}^1_t \|^2 + \mathcal{C}^1_{t+1}(e) 
 \Big\} \Big] & \text{if} \ \ e>0.
\end{cases}
\end{equation}
\end{lemma}

\begin{IEEEproof}
If $e=0$, there is only one feasible scheduling policy:
\begin{equation}
\gamma_t^{\star}(\mathbf{x}_t)=0, \ \ \mathbf{x}_t \in\mathbb{R}^{n}.
\end{equation}
Therefore, 
\begin{equation} \label{eq:vt_0} 
\mathcal{V}_{t}^{\pi}(0) =  \inf_{ \tilde{\mathbf{x}}_t}\ \mathbb{E} \bigg[ \sum_{i\in\{1,2\}}\|X_t^i -  \tilde{x}_t^i\|^2 \bigg] + \mathcal{C}_{t+1}^0(0)  . 
\end{equation}

If $e>0$, the value function in \cref{eq:v_t} can be written as in \cref{eq:vt_alternative}.
\begin{multline}
\mathcal{V}_{t}^{\pi}(e) =  \inf_{ \tilde{\mathbf{x}}_t} \Bigg\{\inf_{\gamma_t}  \int \Bigg[ \Big( \sum_{i\in\{1,2\}}\|x_t^i -  \tilde{x}_t^i\|^2  + \mathcal{C}_{t+1}^0(e) \Big) \mathbb{I}(\gamma_t(\mathbf{x}_t) = 0)  \\ +   \Big( \|x_t^2 -  \tilde{x}_t^2\|^2 + \mathcal{C}_{t+1}^1(e) \Big) \mathbb{I}(\gamma_t(\mathbf{x}_t) = 1) 
 + \Big( \|x_t^1 -  \tilde{x}_t^1\|^2  + \mathcal{C}_{t+1}^1(e) \Big) \mathbb{I}(\gamma_t(\mathbf{x}_t) = 2) \Bigg] \pi(\mathbf{x}_t) d\mathbf{x}_t \Bigg\} \label{eq:vt_alternative}
\end{multline}
For any fixed $\tilde{x}^i_t \in \mathbb{R}^{n_i}$, $i\in\{1,2\}$, the scheduling prescription that achieves the minimum in the inner optimization problem in \cref{eq:vt_alternative} is determined as follows:

\begin{itemize}
\item $\gamma^\star_t(\mathbf{x}_t)=0$ if and only if
\begin{equation}\label{eq:opt_scheduler1}
\|x^i_t -\tilde{x}^i_t \|^2 \leq \mathcal{C}_{t+1}^1(e)-\mathcal{C}_{t+1}^0(e), \ \ i \in \{1,2\};
\end{equation} 

\item $\gamma^\star_t(\mathbf{x}_t)=1$ if and only if
\begin{equation}\label{eq:opt_scheduler2}
\|x^1_t -\tilde{x}^1_t \|^2 > \mathcal{C}_{t+1}^1(e)-\mathcal{C}_{t+1}^0(e)
\end{equation} 
and
\begin{equation}\label{eq:opt_scheduler3}
\|x^1_t -\tilde{x}^1_t \| \geq \|x^2_t -\tilde{x}^2_t \|;
\end{equation} 

\item $\gamma^\star_t(\mathbf{x}_t)=2$ if and only if
\begin{equation}\label{eq:opt_scheduler4}
\|x^2_t -\tilde{x}^2_t \|^2 > \mathcal{C}_{t+1}^1(e)-\mathcal{C}_{t+1}^0(e)
\end{equation} 
and
\begin{equation}\label{eq:opt_scheduler5}
\|x^2_t -\tilde{x}^2_t \| > \|x^1_t -\tilde{x}^1_t \|.
\end{equation} 
\end{itemize} 
Therefore,
\begin{equation}\label{eq:opt_prescription}
\gamma_t^{\star}(\mathbf{x}_t) \Equaldef \begin{cases} 
0,  \ \ \text{if} \ \ \underset{i\in\{1,2\}}{\max} \big\{\|x_t^i - \tilde{x}_t^i \|^2 \big\} \leq \mathcal{C}_{t+1}^1(e)-\mathcal{C}_{t+1}^0(e) \\
\arg \underset{i\in\{1,2\}}{\max}\big\{\|x_t^i - \tilde{x}_t^i \|\big\}, \ \ \text{otherwise}.
\end{cases}
\end{equation}

Using the optimal scheduling prescription in \cref{eq:opt_prescription}, the value function becomes:
\begin{multline}\label{eq:finite_dimensional_cost}
\mathcal{V}_t(e) = \inf_{\tilde{\mathbf{x}}_t} \  \mathbb{E} \Big[ \min\Big\{ \|X^1_t - \tilde{x}^1_t \|^2 + \|X^2_t - \tilde{x}^2_t \|^2 + \mathcal{C}^0_{t+1}(e),\\ \|X^2_t - \tilde{x}^2_t \|^2 + \mathcal{C}^1_{t+1}(e),
\|X^1_t - \tilde{x}^1_t \|^2 + \mathcal{C}^1_{t+1}(e)
 \Big\} \Big].
\end{multline}
\end{IEEEproof}

\cref{lem:finite_dimensional} implies that the optimal solution to \cref{prob:coord} can be found by solving the finite dimensional optimization problem in \cref{eq:finite_dimensional_opt}. 
We will show that \cref{eq:finite_dimensional_opt} admits a globally optimal solution under certain conditions on the probabilistic structure of the problem.



\begin{lemma}\label{lem:solution_finite_opt}
Let $X^1_t$ and $X^2_t$ be independent continuous random vectors with pdfs $\pi_1$ and $\pi_2$. Provided that $\pi_1$ and $\pi_2$ are symmetric and unimodal around $0$, then $\tilde{\mathbf{x}}_t^{\star}=0$ is a global minimizer in \cref{eq:finite_dimensional_opt} for all $e\in\{0,1,\cdots,B\}$.
\end{lemma}

\begin{IEEEproof}
The proof is in Appendix~\ref{ap:analysis}.
\end{IEEEproof}




We are now ready to provide the proof of \cref{thm:main}. 

\begin{IEEEproof}[Proof of Theorem 1] 
We will first show that $(\mathbf{f}^\star,\mathbf{g}^{1\star},\mathbf{g}^{2\star})$ as defined in \cref{thm:main} is globally optimal for \cref{prob:expanded_info}. 

The optimal prescriptions for \cref{prob:coord} are obtained using \Cref{lem:finite_dimensional,lem:solution_finite_opt}. The optimal prescription for the scheduler is given by:
\begin{equation}
\gamma_t^{\star}(\mathbf{x}_t) \Equaldef \begin{cases} 
0,  \ \ \text{if} \ \ \underset{i\in\{1,2\}}{\max} \big\{\|x_t^i \| \big\} < \tau_t^\star(e)\\
\arg \underset{i\in\{1,2\}}{\max}\big\{\|x_t^i \|\big\}, \ \ \text{otherwise},
\end{cases}
\end{equation}
whose threshold functions $\tau^\star_t(e)$ can be computed recursively (see \cref{sec:thresholds}); and the optimal prescription for the estimators are:
\begin{equation}
\tilde{x}_t^{i\star}=0, \ \ i\in\{1,2\}.
\end{equation}
Therefore, using the equivalence between \cref{prob:expanded_info} and \cref{prob:coord}, the optimal strategy profiles for \cref{prob:expanded_info} are
\begin{equation}
f^\star_t(\mathbf{x}_t,e_t) \Equaldef \begin{cases} 
0,  \ \  \text{if} \ \ \underset{i\in\{1,2\}}{\max} \big\{\|x_t^i \| \big\} < \tau^\star_t(e_t) \\
\arg \underset{i\in\{1,2\}}{\max}  \|x^i_t\|, \ \ \ \ \   \text{otherwise,}
\end{cases}
\end{equation}
and
\begin{equation}
g^{i\star}_t\big(y_t^i) \Equaldef \begin{cases} 
x_t^i & \textup{if} \ \ y_t^i = x_t^i \\
0 & \textup{if} \ \ y_t^i = \varnothing
\end{cases},
\ \  i\in\{1,2\}.
\end{equation}

Moreover, since the solution to \cref{prob:expanded_info}, $(\mathbf{f}^\star,\mathbf{g}^{1\star},\mathbf{g}^{2\star})$  does not depend on the additional information provided to the estimators and is adapted to the original information structure of the estimators in \cref{prob:main}, it is also a globally optimal strategy profile for \cref{prob:main}.



\end{IEEEproof}





\section{Computation of optimal thresholds}\label{sec:thresholds}
Once the structural result in \cref{thm:main} is established, the optimal scheduling strategy is completely specified by the sequence of optimal threshold functions $\tau^\star_t$, $t\in\{1,\cdots,T\}$. The thresholds $\tau^\star_t(e)$ are obtained using the functions $\mathcal{C}_{t+1}^{0}(e),\mathcal{C}_{t+1}^{1}(e)$ in \cref{eq:auxiliary1,eq:auxiliary2}. The functions $\mathcal{C}_t^0(\cdot),\mathcal{C}_t^1(\cdot)$ can be computed by computing the value functions $\mathcal{V}^{\pi}_t$ via a backward inductive procedure. Note that we can simplify the expression for the value function using \Cref{lem:solution_finite_opt} and \cref{eq:finite_dimensional_opt} to:
\begin{equation}
\mathcal{V}^{\pi}_t(0) =  \mathbb{E}\Big[\|X^1_t\|^2 + \|X^2_t\|^2 +  \mathcal{V}^{\pi}_{t+1}\big(\min\{Z_t,B\}\big) \Big]  \label{eq:v_0_final}
\end{equation}
and
\begin{equation}
\mathcal{V}^{\pi}_t(e) = \mathbb{E}\Big[ \min\Big\{\|X^1_t\|^2 + \|X^2_t\|^2 + \mathcal{C}^0_{t+1}(e),  \\
\|X^2_t\|^2+\mathcal{C}^1_{t+1}(e),  \|X^1_t\|^2+\mathcal{C}^1_{t+1}(e) \Big\} \Big] \ \ \text{if} \ \ e>0. \label{eq:v_final}
\end{equation}

The following algorithm outlines the recursive computation of the threshold function $\tau_t^\star$:

\begin{algorithm}[H]
	\caption{Computing the optimal threshold functions $\tau^\star_t$}
	\begin{algorithmic}
		\STATE Initialization: 
    \STATE $t \leftarrow T$

		\STATE Set $\mathcal{V}_{T+1}^{\pi}(e) \gets 0$ for $e\in\{0,\cdots,B\}$
		\WHILE{ $t \geq 1$}
    \STATE Compute $\mathcal{C}_{t+1}^{0}(e)$ and $\mathcal{C}_{t+1}^{1}(e)$ using  \cref{eq:auxiliary1,eq:auxiliary2} for $e\in\{1,\cdots,B\}$  
		\STATE Set $\tau^\star_t(e) \gets  \sqrt{\mathcal{C}^1_{t+1}(e)-\mathcal{C}^0_{t+1}(e)}$ for $e\in\{1,\cdots,B\}$ 
		\STATE Compute $\mathcal{V}_t^{\pi}(e)$ using \ \cref{eq:v_0_final,eq:v_final} for $e\in\{0,\cdots,B\}$
		
		\STATE{$t  \leftarrow t-1$}
		\ENDWHILE
	\end{algorithmic}
\end{algorithm}

\begin{remark}
The expectations in the algorithm are with respect to the random vectors $X^1_t$ and $X^2_t$. Computing these expectations for high dimensional random vectors may be computationally intensive for some source distributions, but in practice they can be approximated using Monte Carlo methods. The remaining operations in the algorithm admit efficient implementations.
\end{remark}

\section{Illustrative examples}

\subsection{Optimal blind scheduling}

Before we provide a few numerical examples it is useful to introduce a scheduling strategy which is based exclusively on the statistics of the sources, and not on the observations. Consider the following \textit{blind} scheduling strategy: if the battery is not empty, transmit the source whose variance is the largest, i.e.,
\begin{equation}\label{eq:blind_strategy}
f_t^{\mathrm{blind}} (e_t)\Equaldef \begin{cases} 0 & \text{if} \ e_t=0 \\
\arg \underset{i\in\{1,2\}}{\max} \Big\{ \mathbb{E}\big[\|X^i_t - \mathbb{E}[X^i_t]\|^2\big] \Big\} & \text{otherwise},
\end{cases}
\end{equation}
The estimation strategies associated with blind scheduling are:
\begin{equation}
g^{\mathrm{blind}\ i}_t\big(y_t^i) \Equaldef \begin{cases} 
x_t^i & \textup{if} \ \ y_t^i = x_t^i \\
\mathbb{E}[X^i_t] & \textup{if} \ \ y^i_t = \varnothing
\end{cases},
\ \  i\in\{1,2\}.
\end{equation}

The performance of the blind scheduling and estimation strategies and the  is given by:
\begin{multline}\label{eq:blind}
\mathcal{J}^{\mathrm{blind}}(B)\Equaldef\sum_{t=1}^T\Big[ \mathbb{P}(E_t=0)\sum_{i\in\{1,2\}}\mathbb{E}\big[\|X^i_t - \mathbb{E}[X^i_t]\|^2\big]\\
+\big(1-\mathbb{P}(E_t=0)\big)\min_{i\in\{1,2\}} \Big\{ \mathbb{E}\big[\|X^i_t-\mathbb{E}[X^i_t]\|^2\big] \Big\}\Big],
\end{multline}
where the probabilities $\big\{\mathbb{P}(E_t=0)$, $t\in\{1,\cdots,T\}\big\}$ are computed recursively using \cref{eq:c2,eq:battery} and assuming $E_1=B>0$ with probability $1$.

\begin{example}[Limited number of transmissions]\label{ex:limited}
Consider the scheduling of two i.i.d. zero mean scalar Gaussian sources with variances $\sigma_1^2 = \sigma_2^2 =1$. Assume that the total system deployment time is $T$, and that during that time the scheduler is only allowed to transmit $B<T$ times. Furthermore, assume that during that time, there is no energy being harvested, i.e., $Z_t=0$ with probability $1$, and there are no additional communication costs.


The algorithm outlined in \cref{sec:thresholds} is used to compute the optimal thresholds, which are functions of the time index, and the energy level at the battery. \Cref{fig:thresholds} displays the optimal thresholds computed for this example with $T=100$ and $B=30$.

\begin{figure}[t!]
\begin{center}
\includegraphics[width=0.5\textwidth]{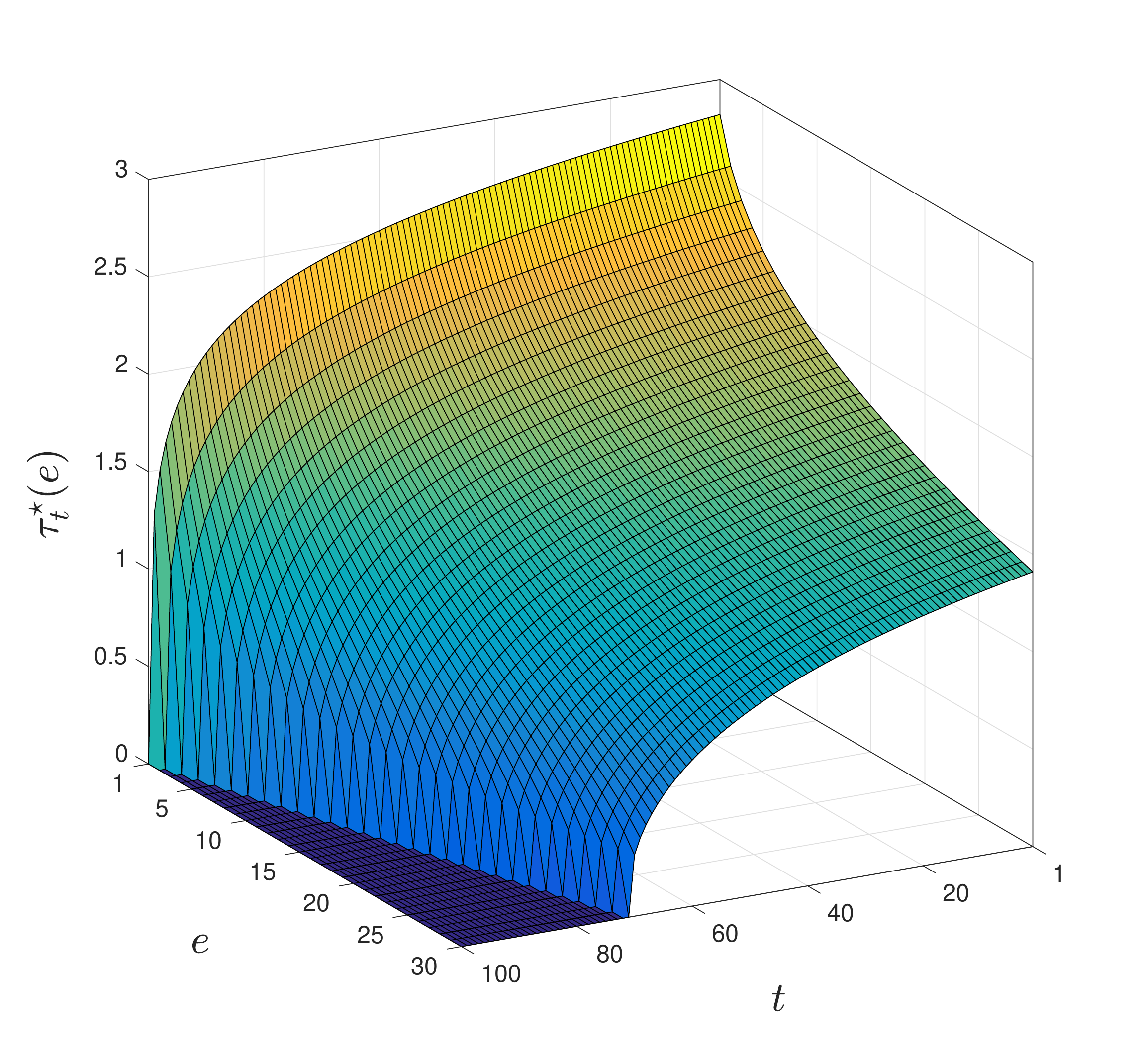}    
\caption{Optimal threshold function for the scheduling of two i.i.d. standard Gaussian sources. The threshold is a function of the energy level and time.}  
\label{fig:thresholds}                                 
\end{center}                                 
\end{figure}

Notice that when the energy level is greater than the remaining deployment time, the optimal threshold is zero, that is, the observation with the largest magnitude is always transmitted. On the other hand, if the power level is below the remaining deployment time, the optimal threshold is strictly positive and it increases as the power level decreases. That means that as the battery depletes, the scheduler will only transmit observations whose magnitudes are increasingly larger.
\end{example}

\begin{figure}[t!]
\begin{center}
\includegraphics[width=0.5\textwidth]{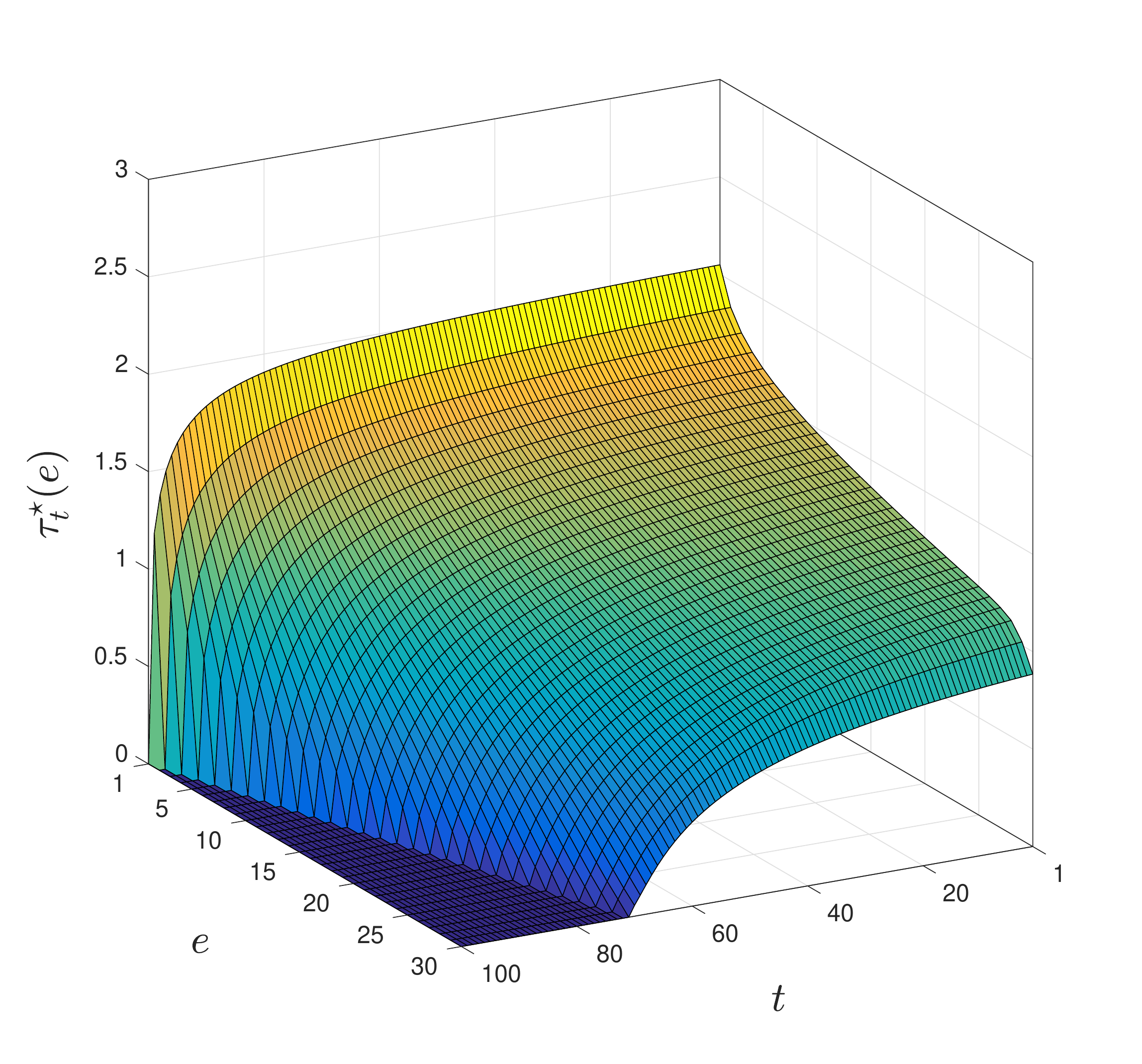}
\caption{Optimal threshold function for the scheduling of two i.i.d. standard Gaussian sources with energy harvesting. The threshold is a function of the energy level and time.} 
\label{fig:thresholds_harvesting}                           
\end{center}                               
\end{figure}

\begin{example}[Energy harvesting scheduler] Consider a setup identical to that in \cref{ex:limited}, but in addition assume that the energy harvesting process $Z_t$ is distributed according to two possible probability mass functions:
\begin{equation}\label{eq:harvesting}
p^1_Z(z) = \begin{cases}
0.85 & z=0\\
0.1 & z=1\\
0.05 & z=2
\end{cases}
\ \ \text{or} \ \ p^2_Z(z) = \begin{cases}
0.7 & z=0\\
0.2 & z=1\\
0.1 & z=2
\end{cases},
\end{equation}
yielding on average $0.2$ and $0.4$ energy units per time step, respectively.

The optimal thresholds obtained for the energy harvesting system under $p^1_Z$ are shown in \cref{fig:thresholds_harvesting}, and they are uniformly smaller than the ones of the system without harvesting. We also note a change in the ``curvature'' of the threshold function for a fixed $t$.
\end{example}

\begin{figure}[ht!]
\begin{center}
\includegraphics[width=0.5\textwidth]{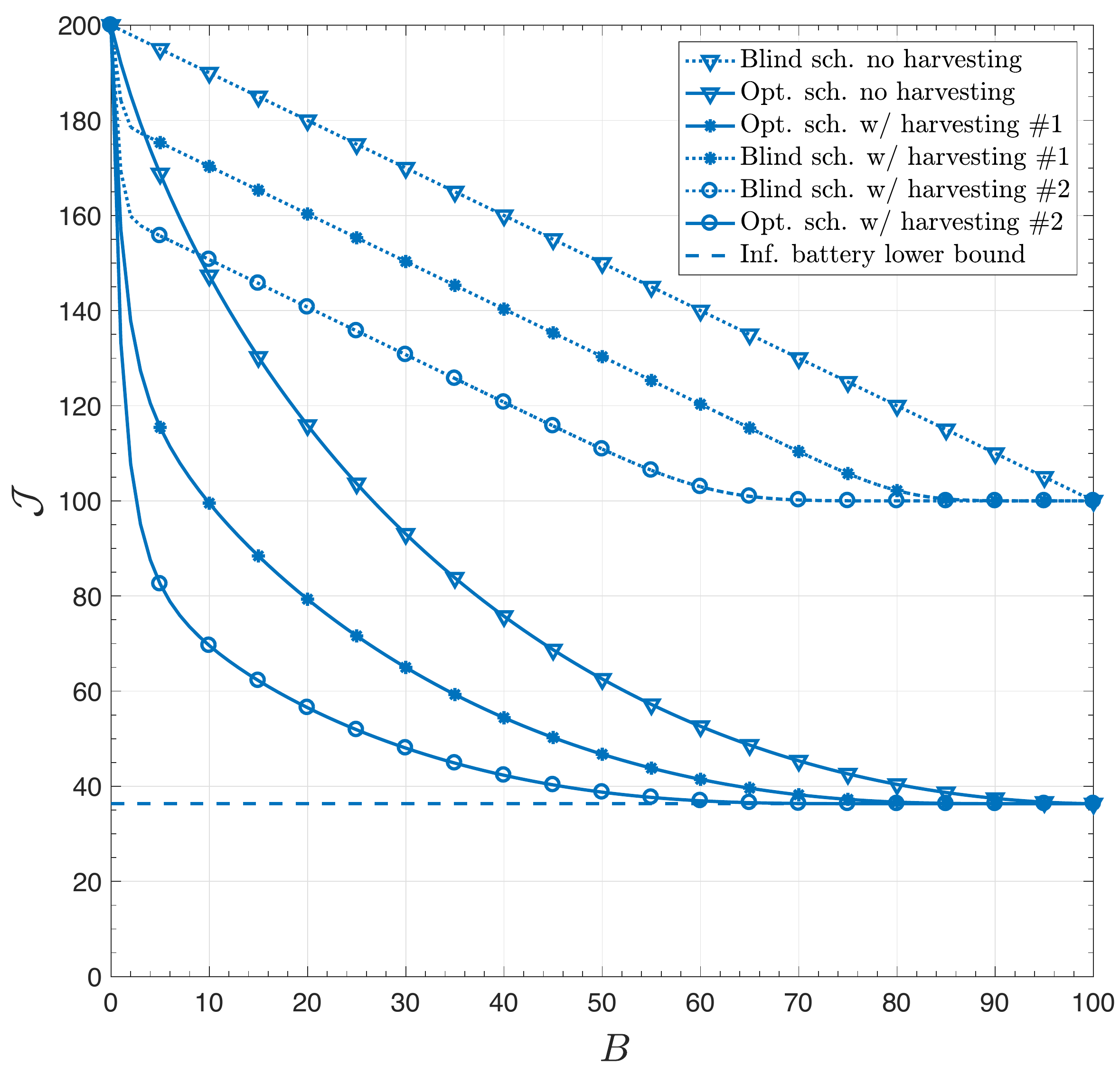}    
\caption{Comparison between the performances of the optimal open-loop and closed-loop strategies as a function of the battery capacity, $B$. The relative gap between these two curves is defined as the Value of Information.}  
\label{fig:performance}                                 
\end{center}                                 
\end{figure}

\begin{figure}[ht!]
\begin{center}
\includegraphics[width=0.5\textwidth]{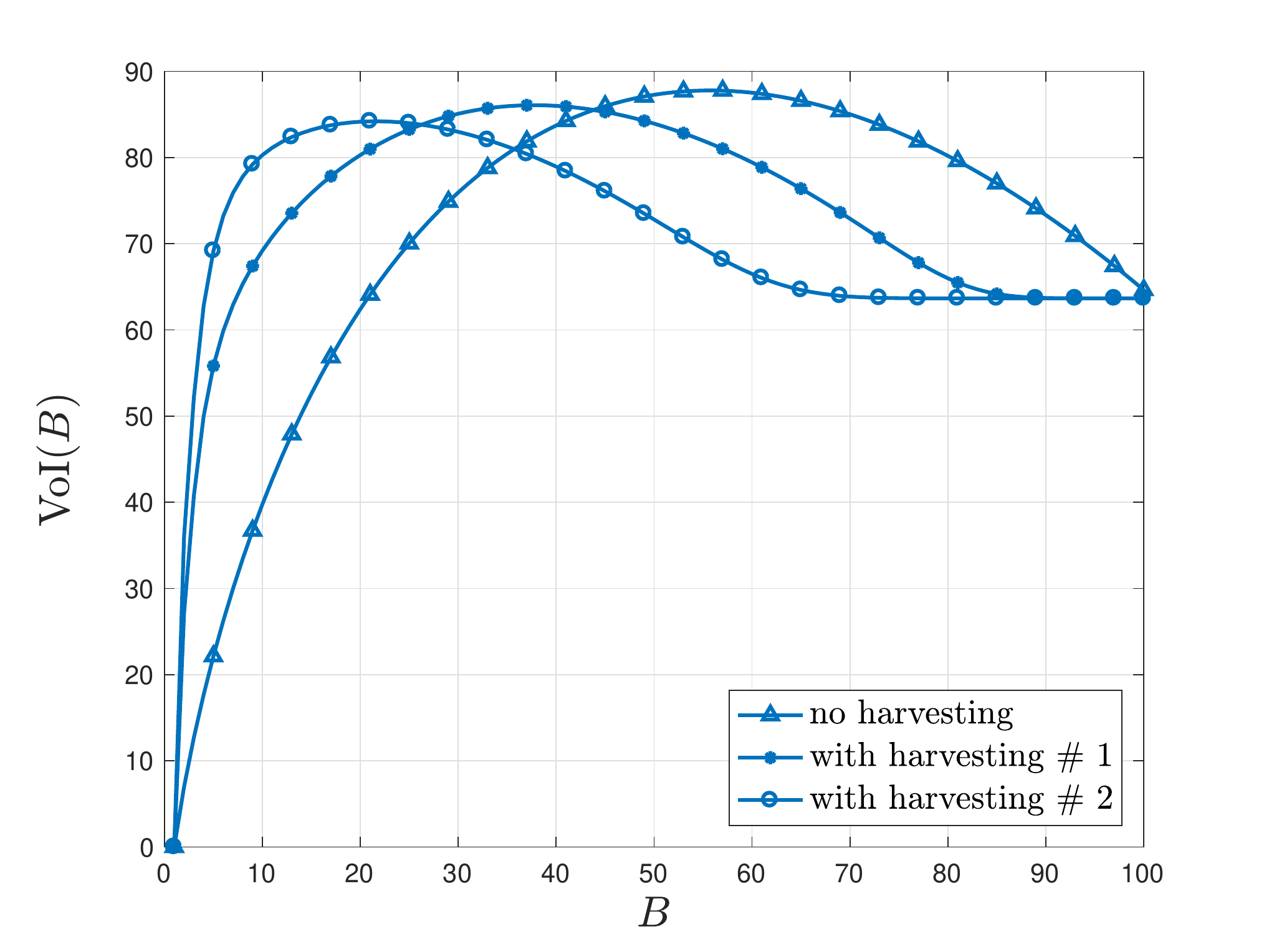}    
\caption{Value of Information in the numerical example as a function of the battery capacity.}  
\label{voi}                                 
\end{center}                                 
\end{figure}

 \Cref{fig:performance} shows the performance of the optimal strategy and the blind scheduling scheme as a function of the battery capacity $B$ for the three systems: no harvesting, harvesting with $p_Z^1$ and $p_Z^2$. The optimal scheme proposed in this paper leads to a significant improvement upon the blind scheduling strategy of \cref{eq:blind_strategy}. The gap in performance between optimal open-loop and closed-loop strategies is defined as the ``Value of information'' ($\mathrm{VoI}$). Mathematically, the VoI is given by:
\begin{equation}
\mathrm{VoI}(B) \Equaldef \mathcal{J}^{\mathrm{blind}}(B)-\mathcal{J}^{\star}(B).
\end{equation}

\Cref{voi} illustrates the VoI as a function of the battery capacity $B$. Notice that this function is not monotonic increasing in the battery capacity. Therefore, there exists a nontrivial optimal value $B^\star$ such that the $\mathrm{VoI}$ is maximized. In \cref{ex:limited}, the value of $B$ that maximizes the VoI in the system without energy harvesting is $B^\star=55$.

For $B=10$, without energy harvesting, the optimal performance is $\mathcal{J}^{\star} \approx 147.37$. However, in order to achieve a comparable performance using blind scheduling, a battery of capacity equal to $53$ energy units would be required. Therefore, the energy savings in this case is of approximately $81.13\%$.



\section{Extensions}

\subsection{The $N$ sensor case}\label{sec:N_sensors}

\cref{thm:main} holds for any number of sensors ($N\geq2$). Let $\mathbf{x}_t=(x_t^1,x_t^2,\cdots,x_t^N)$, where $x_t^i \in \mathbb{R}^{n_i}$ is the observation at the $i$-th sensor. Provided that the observations are mutually independent and, their pdfs are symmetric and unimodal around $a^1,a^2,\cdots,a^N$, where $a_i\in\mathbb{R}^{n_i}$, $i\in \{1,2,\cdots, N\}$, the jointly optimal scheduling and estimation strategies are:
\begin{equation}
f_t^{\star}(\mathbf{x}_t,e_t) \Equaldef \begin{cases} 
0,  \ \ \text{if} \ \ \underset{i\in\{1, \cdots, N\}}{\max} \big\{\|x_t^i - a^i \| \big\} \leq \tau_t^\star(e_t)\\
\arg \underset{i\in\{1,\cdots,N\}}{\max}\big\{\|x_t^i -a^i\|\big\}, \ \ \text{otherwise},
\end{cases}
\end{equation}
and
\begin{equation}
g^{i\star}_t\big(y^i) \Equaldef \begin{cases} 
x_t^i & \textup{if} \ \ y_t^i = x_t^i \\
a^i  & \textup{if} \ \ y_t^i = \varnothing
\end{cases}, \ \ i\in\{1,\cdots,N\}.
\end{equation}

\subsection{Unequal weights and communication costs}

In certain applications each sensor may be assigned a different weight in the expected distortion metric. This is done to emphasize the importance of the observations made by one sensor relative to another. Additionally, different sensors may also have different communication costs, which may reflect the dimension of the measurements or used to preserve the battery power, for instance. These cases are captured by the following cost functional:
\begin{equation}\label{eq:cost_general}
\mathcal{J}\big(\mathbf{f},\mathbf{g}^1,\mathbf{g}^2\big) \Equaldef 
\sum_{t=1}^T \mathbb{E}\Bigg[  \sum_{i\in\{1,2\}}w_i\|X^i_t-\hat{X}^i_t \|^2  +c_i\mathbb{I}(U_t = i) \Bigg].
\end{equation}

\begin{figure}[ht!]
\begin{center}
\includegraphics[scale=0.45]{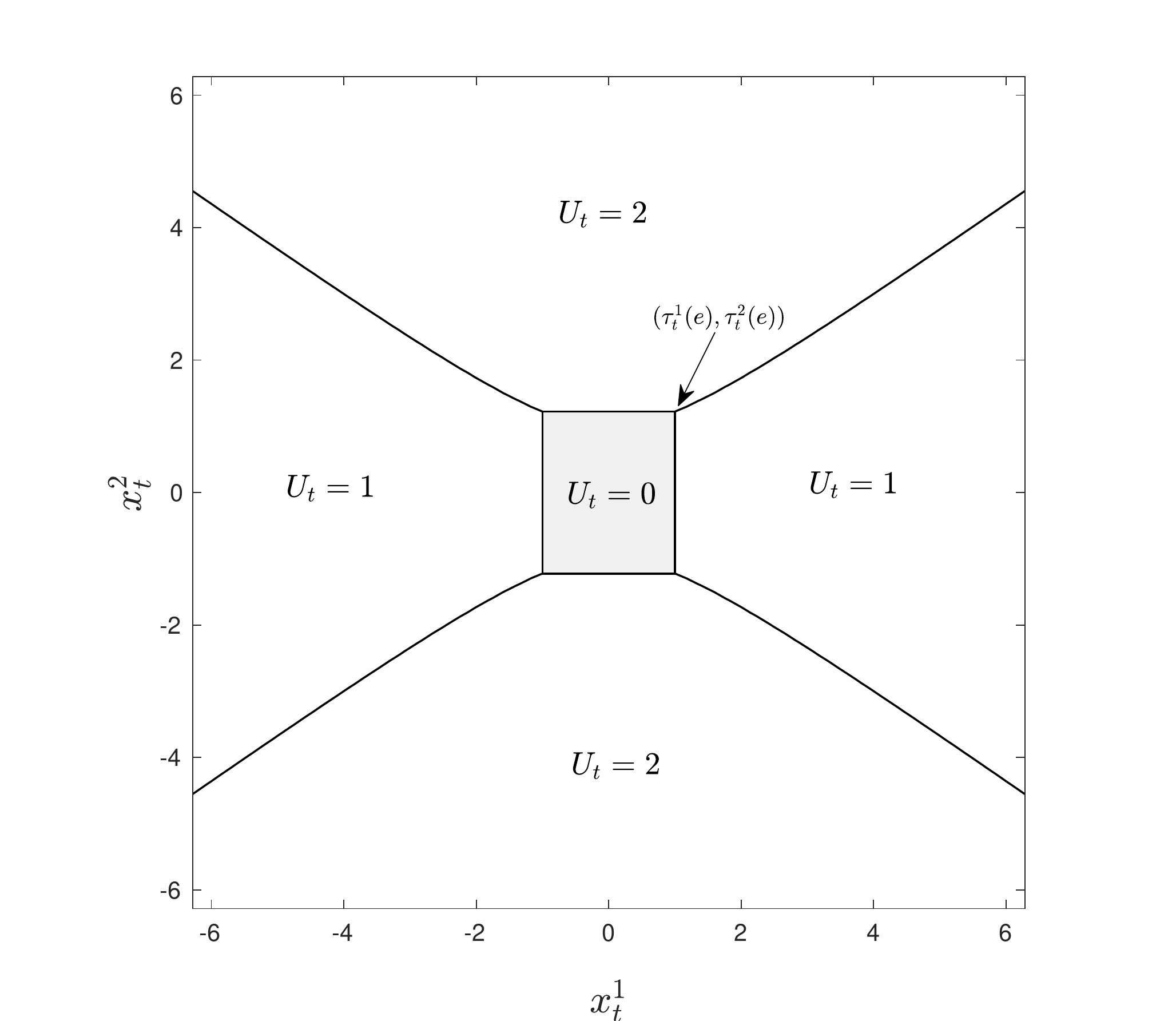}    
\caption{Optimal scheduling decision region for a cost function with unequal weights and communication costs. Notice that the boundaries delimiting the scheduling decisions corresponding to $U_t=1$ and $U_t=2$ are not straight lines. }
\label{fig:unequal}                                 
\end{center}                                 
\end{figure}

The globally optimal scheduling and estimation strategies for the more general cost functional in \cref{eq:cost_general} are given by \cref{eq:scheduling_general,eq:estimation_general}. In order to illustrate the main difference from the case with uniform weights, consider \cref{fig:unequal}, which shows a ``no-transmission region'' characterized by a rectangle defined by threshold functions $(\tau^1_t(e),\tau^2_t(e))$ and a hyperbola that separates the regions associated with scheduling sensors 1 and 2. In contrast, the ``no-transmission  region'' for the uniform case is characterized by a square defined by a single threshold $\tau_t(e)$ and a hyperplane which separates the transmission regions for sensors 1 and 2.

\begin{figure*}
\begin{equation}\label{eq:scheduling_general}
f^\star_t(\mathbf{x}_t,e_t)=\begin{cases}
0, & \ \text{if}\ \|x_t^1 - a^1\| \leq \sqrt{\frac{\tau^1_t(e_t)}{w_1}}, \ \|x_t^2- a^2\| \leq \sqrt{\frac{\tau^2_t(e_t)}{w_2}} \\
1, & \ \text{if}\ \|x_t^1- a^1\| > \sqrt{\frac{\tau^1_t(e)}{w_1}}, \ w_1\|x_t^1- a^1\|^2 - w_2\|x_t^2- a^2\|^2 \geq \tau^1_t(e_t)-\tau^2_t(e_t) \\
2, & \text{otherwise}
\end{cases}
\end{equation}
\begin{equation}\label{eq:estimation_general}
g^{i\star}_t\big(y_t^i) \Equaldef \begin{cases} 
x_t^i & \textup{if} \ \ y_t^i = x_t^i \\
a^i  & \textup{if} \ \ y_t^i = \varnothing
\end{cases}, \ \ i\in\{1,2\}
\end{equation}
\hrulefill
\end{figure*}

\section{Conclusions}
This paper studies the problem of optimal scheduling in a sequential remote estimation system where non-collocated sensors and estimators communicate over a shared medium. The access to the communication resources is granted by an energy-harvesting scheduler, which implements an observation-driven medium access control scheme in order to avoid packet collisions. The underlying assumption is that the sensors make measurements that are independent and identically distributed in time, but the energy level at the scheduler has a  stochastic dynamics, which couples the decision-making process in time. The optimal solutions to such remote estimation problems are typically very difficult to find due to the presence of signaling between the scheduler and estimators.

The main result herein is to establish, under certain assumptions on the probabilistic model of the sources, the joint optimality of a pair of scheduling and estimation strategies. More importantly, the globally optimal solution is obtained in spite of the lack of convexity in the objective function introduced by signaling. The overarching proof consists of a judicious expansion of the information sets at the estimators, which enables the use of the common information approach to solve a single dynamic program from the perspective of a fictitious coordinator. Finally, by noticing that the optimal solution to this ``relaxed'' problem does not depend on the additional information introduced in the expansion, it is also shown to be optimal for the original optimization problem. As a byproduct, our proof technique also applies to more general settings with an arbitrary number of sensors, unequal weights and communication costs. 

Future work in this problem includes the scheduling of correlated sources, but independent in time; independent Gauss-Markov sources (some progress in this area was reported in \cite{Gagrani:2018}); and networks prone to packet-drops.


\appendices

\section{Auxiliary results}
\label{Ap:Cont}

The following two definitions and theorem can be found in \cite{Burchard:2009} and in \cite{Hardy:1954}.

\begin{definition}[Symmetric rearrangement]
Let $\mathbb{A}$ be a measurable set of finite volume in $\mathbb{R}^n$. Its symmetric rearrangement $\mathbb{A}^*$ is defined as the open ball centered at $0 \in \mathbb{R}^n$ whose volume agrees with $\mathbb{A}$.
\end{definition}

\begin{definition}[Symmetric decreasing rearrangement]
Let $f: \mathbb{R}^n \rightarrow \mathbb{R}$ be a nonnegative measurable function that vanishes at infinity. The symmetric decreasing rearrangement $f^{\downarrow}$ of $f$ is
\begin{equation}
f^{\downarrow} (x) \Equaldef \int_{0}^{\infty} \mathbb{I}\Big(x\in {\{\xi\in \mathbb{R}^n \mid f(\xi)>t\}^*}\Big)dt.
\end{equation}
\end{definition}

\begin{theorem}[Hardy-Littlewood Inequality]
If $f$ and $g$ are two nonnegative measurable functions defined on $\mathbb{R}^n$ which vanish at infinity, then the following holds:
\begin{equation}
\int_{\mathbb{R}^n}f(x)g(x)dx\leq \int_{\mathbb{R}^n}f^{\downarrow}(x)g^{\downarrow}(x)dx,
\end{equation}
where $f^{\downarrow}$ and $g^{\downarrow}$ are the symmetric decreasing rearrangements of $f$ and $g$, respectively.
\end{theorem}

\section{Proof of \cref{lem:solution_finite_opt}}\label{ap:analysis}

\subsection{Empty battery}
Let $e=0$. The value function in \cref{eq:finite_dimensional_opt} is given by
\begin{equation}
\mathcal{V}_{t}^{\pi}(0) =
\inf_{ \tilde{\mathbf{x}}_t} \ \mathbb{E} \Big[ \sum_{i\in\{1,2\}}\|X_t^i -  \tilde{x}_t^i\|^2 \Big] + \mathcal{C}_{t+1}^0(0).
\end{equation}
The infimum in the expression above is achieved by 
\begin{equation}
\tilde{\mathbf{x}}^{\star}_t = \big( \mathbb{E}[X_t^1],\mathbb{E}[X_t^2]\big).
\end{equation}
Since $\pi_1$ and $\pi_2$ are symmetric around $0$,
\begin{equation}
\tilde{\mathbf{x}}^{\star}_t = 0.
\end{equation}
Therefore, if $e=0$, the infimum in \cref{eq:finite_dimensional_opt} is achieved by:
\begin{align}
\tilde{\mathbf{x}}_t^{\star} &= 0, \ \ i\in\{1,2\}.
\end{align}

\subsection{Nonempty battery} 
Let $e>0$. The value function in \cref{eq:finite_dimensional_opt} is given by
\begin{multline}\label{vt_min}
\mathcal{V}_{t}^{\pi}(e) = \inf_{ \tilde{\mathbf{x}}_t} \  \mathbb{E} \Big[ \min \Big\{  \|X_t^1 -  \tilde{x}_t^1\|^2 + \|X_t^2 -  \tilde{x}_t^2\|^2 + \mathcal{C}_{t+1}^0(e) , \\
  \|X_t^2 -  \tilde{x}_t^2\|^2 + \mathcal{C}_{t+1}^1(e) ,   \|X_t^1 -  \tilde{x}_t^1\|^2  + \mathcal{C}_{t+1}^1(e)  \Big\}   \Big].
\end{multline}

The function $\mathcal{C}^1_{t+1}(e)$ is always larger than $\mathcal{C}^0_{t+1}(e)$ for all $e\in\{1,\cdots,B\}$ and $t\in\{1,\cdots,T\}$. In order to establish this fact, notice that: 
\begin{align*}
\mathcal{C}^1_{t+1}(e) & = c + \mathcal{C}^0_{t+1}(e-1) \\
& \stackrel{(a)}{\geq} \mathcal{C}^0_{t+1}(e-1) \\
& \stackrel{(b)}{\geq} \mathcal{C}^0_{t+1}(e),
\end{align*}
where the inequality $(a)$ follows from the fact that $c\geq0$; and inequality $(b)$ follows from the value functions $\mathcal{V}^{\pi}_t(e)$ being non-increasing in $e$. This can be argued using the fact that, at any time step, having more energy available for transmission cannot reduce the optimal performance of the system.

The optimization problem in \cref{vt_min} is equivalent to:
\begin{equation}
\inf_{ \tilde{\mathbf{x}}_t} \  \mathbb{E} \Big[ \min \Big\{  \|X_t^1 -  \tilde{x}_t^1\|^2 + \|X_t^2 -  \tilde{x}_t^2\|^2  , \\
  \|X_t^2 -  \tilde{x}_t^2\|^2 + \kappa_t(e) ,   \|X_t^1 -  \tilde{x}_t^1\|^2  + \kappa_t(e) \Big\}   \Big],
\end{equation}
where
\begin{equation}
\kappa_t(e)\Equaldef \mathcal{C}_{t+1}^1(e)- \mathcal{C}_{t+1}^0(e).
\end{equation}

Consider the auxiliary cost function $\mathcal{J}^{e}_t:\mathbb{R}^{n_1}\times\mathbb{R}^{n_2} \rightarrow \mathbb{R}$ defined as
\begin{equation}
\mathcal{J}^{e}_t(\tilde{\mathbf{x}}_t) \Equaldef \mathbb{E} \Big[\min\big\{\|X^1_t -\tilde{x}^1_t \|^2 + \|X^2_t -\tilde{x}^2_t \|^2, \\ \|X^2_t -\tilde{x}^2_t \|^2 + \kappa_t(e),
\|X^1_t -\tilde{x}^1_t \|^2 + \kappa_t(e) \big\} \Big],
\end{equation}
where the expectation is taken with respect to the random vectors $X^1_t$ and $X^2_t$.

The remainder of the proof consists of solving the following optimization problem: 
\begin{equation}
\underset{\tilde{\mathbf{x}}_t}{\inf} \ \ \mathcal{J}_t^e(\tilde{\mathbf{x}}_t).
\end{equation}
Define the function $\mathcal{G}:\mathbb{R}^n \times\mathbb{R}^n \rightarrow \mathbb{R}$ such that
\begin{equation}\label{eq:G}
\mathcal{G}_t^e(\tilde{\mathbf{x}}_t;\mathbf{x}_t) \Equaldef \min\big\{\|x_t^1-\tilde{x}_t^1\|^2 + \|x_t^2-\tilde{x}_t^2\|^2,\\ \|x_t^2-\tilde{x}_t^2\|^2 + \kappa_t(e), \|x_t^1-\tilde{x}_t^1\|^2+\kappa_t(e) \big\}.
\end{equation}
Using the fact that $X_t^1$ and $X_t^2$ are independent, and the function $\mathcal{G}^e_t$ defined in \cref{eq:G}, we can rewrite the function $\mathcal{J}^e_t(\tilde{\mathbf{x}}_t)$ in integral form as:
\begin{equation}
\mathcal{J}^e_t(\tilde{\mathbf{x}}_t) = \int_{\mathbb{R}^{n_2}} \Bigg[\int_{\mathbb{R}^{n_1}} \mathcal{G}^e_t(\tilde{\mathbf{x}}_t;
\mathbf{x}_t) \pi_1(x_t^1)dx_t^1 \Bigg]\pi_2(x_t^2)dx_t^2.
\end{equation}

\begin{figure}[!t]
\centering

\begin{tikzpicture}[domain=-10:10,scale=6,samples=200]
    {}

\draw[thick,dotted] (0.6,1/5) -- (1,1/5);

\draw[thick,dotted] (0.6,1/5) -- (0.6,0);

 \draw[thick,dotted] (0.1,0.5) -- (1.1,0.5) node[above] {$\|x^2_t-\tilde{x}^2_t\|^2+\kappa_t(e)$}; 

\draw[very thick,scale=1,domain=0.2:1,smooth,variable=\x,color=black]  plot ({\x},{min((2*\x-1.2)*(2*\x-1.2)+1/5,0.5)});

\draw[->, thick] (0.1,0) -- (1.2,0) node[right] {$x^1_t$};
   
\foreach \x in {0.6} \draw (\x cm,0.25pt) -- (\x cm,-0.25pt);

\draw (0.6,-0.02)  node[below] {$\tilde{x}^1_t$};

\draw (1.1,1/5)  node[above] {$\min\big\{\kappa_t(e),\|x^2_t-\tilde{x}^2_t\|^2\big\}$};

\draw (0.25,0.6)  node[above] {\textcolor{black}{$\mathcal{G}^e_t(\tilde{\mathbf{x}}_t;\mathbf{x}_t)$}};

\end{tikzpicture}

\caption{Conceptual plot of $\mathcal{G}^e_t(\tilde{\mathbf{x}}_t;\mathbf{x}_t)$ as a function of $x^1_t$ while keeping its remaining arguments fixed. This figure illustrates that when the norm of $\tilde{x}^1_t$ tends to infinity the function remains bounded.} 
\label{fig:property}
\end{figure}
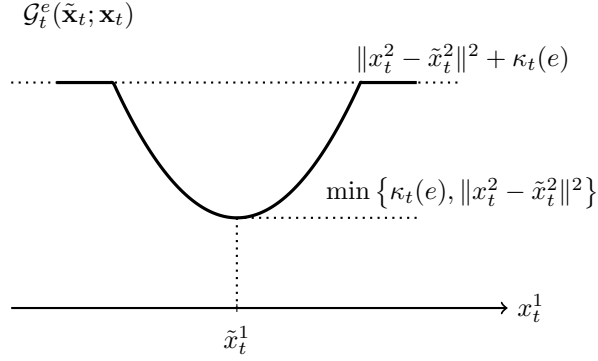

The function $\mathcal{G}^e_t$ can be alternatively represented as:
\begin{equation}\label{eq:Ge}
\mathcal{G}^e_t(\tilde{\mathbf{x}}_t;\mathbf{x}_t) = \min\Big\{\|x_t^2-\tilde{x}_t^2\|^2 + \kappa_t(e), \\
\|x_t^1-\tilde{x}_t^1\|^2 + \min\big\{\kappa_t(e), \|x_t^2-\tilde{x}_t^2\|^2\big\} \Big\}.
\end{equation}
The function in \cref{eq:Ge} is sketched in \cref{fig:property} as a function of $x_t^1$ while keeping $x_t^2$ and $\tilde{\mathbf{x}}_t$ fixed. 

Finally, let the function $\mathcal{H}^e_t:\mathbb{R}^n \times \mathbb{R}^n \rightarrow \mathbb{R}$ be defined as:
\begin{equation}
\mathcal{H}^e_t(\tilde{\mathbf{x}}_t;\mathbf{x}_t) \Equaldef \|x_t^2-\tilde{x}_t^2\|^2 + \kappa_t(e) -\mathcal{G}^e_t(\tilde{\mathbf{x}}_t;\mathbf{x}_t).
\end{equation}

Notice that the function $\mathcal{H}^e_t$ vanishes as the norm of $x^1_t$ tends to infinity, i.e.,
\begin{equation}
\lim_{\|x^1_t\|\rightarrow +\infty} \mathcal{H}^e_t(\tilde{\mathbf{x}}_t;\mathbf{x}_t) = 0.
\end{equation}
From the Hardy-Littlewood inequality (see Appendix~\ref{Ap:Cont}), we have:
\begin{equation}
\int_{\mathbb{R}^{n_1}}\mathcal{H}^e_t(\tilde{\mathbf{x}}_t;\mathbf{x}_t)\pi_1(x_t^1)dx_t^1  \leq  \int_{\mathbb{R}^{n_1}}\mathcal{H}^{e\downarrow}_t(\tilde{\mathbf{x}}_t;\mathbf{x}_t)\pi_1^{\downarrow}(x^1_t)dx^1_t,
\end{equation}
where $\pi_1^{\downarrow}$ and $\mathcal{H}^{e\downarrow}_t$ denote the symmetric decreasing rearrangements  of $\pi_1$ and $\mathcal{H}^e_t$, respectively. The following facts hold:

\begin{enumerate} 
\item Since ${\pi_1}$ is symmetric and unimodal around $0$,
\begin{equation}
\pi_1^{\downarrow} = \pi_1.
\end{equation}

\item Since $\mathcal{H}^e_t(\tilde{\mathbf{x}}_t;\mathbf{x}_t)$, as a function of $x^1_t$, is symmetric and unimodal around $\tilde{x}^1_t$ (a fact that can be verified by inspection), we have:
\begin{equation}
\mathcal{H}^{e\downarrow}_t(\tilde{\mathbf{x}}_t;\mathbf{x}_t) = \mathcal{H}^e_t\big((0,\tilde{x}^2_t);\mathbf{x}_t\big).
\end{equation}
\end{enumerate}

Therefore, the Hardy-Littlewood inequality implies that:
\begin{equation}
\int_{\mathbb{R}^{n_1}}\mathcal{H}^e_t(\tilde{\mathbf{x}}_t;\mathbf{x}_t)\pi_1(x^1_t)dx_t^1  \leq  \int_{\mathbb{R}^{n_1}}\mathcal{H}^e_t\big((0,\tilde{x}^2_t);\mathbf{x}_t\big)\pi_1(x_t^1)dx_t^1,
\end{equation}
which is equivalent to:
\begin{equation}
\int_{\mathbb{R}^{n_1}}\mathcal{G}^e_t\big((0,\tilde{x}^2_t);\mathbf{x}_t\big)\pi_1(x_t^1)dx_t^1 \leq \int_{\mathbb{R}^{n_1}}\mathcal{G}^e_t(\tilde{\mathbf{x}}_t;\mathbf{x}_t)\pi_1(x_t^1)dx_t^1.
\end{equation}
Therefore, 
\begin{equation} 
\tilde{x}_t^{1\star}=0.
\end{equation}

Fixing $\tilde{x}^{1\star}_t=0$ and following the same sequence of arguments exchanging the roles of $x^1_t$ and $x^2_t$, we show that $\tilde{x}^{2\star}_t=0$. Therefore, 
\begin{equation}
\tilde{\mathbf{x}}_t^{\star} = 0.
\end{equation}

\bibliographystyle{IEEEtran}        
\bibliography{Scheduling}

\end{document}